\documentclass[pdflatex,sn-mathphys-num]{sn-jnl}

\usepackage{graphicx}%
\usepackage{multirow}%
\usepackage{amsmath,amssymb,amsfonts}%
\usepackage{amsthm}%
\usepackage{mathrsfs}%
\usepackage[title]{appendix}%
\usepackage{xcolor}%
\usepackage{textcomp}%
\usepackage{manyfoot}%
\usepackage{booktabs}%
\usepackage{algorithm}%
\usepackage{algorithmicx}%
\usepackage{algpseudocode}%
\usepackage{listings}%
\usepackage{subcaption}
\usepackage{tabularx}
\usepackage{CJKutf8}
\usepackage{url}
\newcommand*{\Ja}[1]{%
  \begin{CJK}{UTF8}{ipxm}#1\end{CJK}}
  
\theoremstyle{thmstyleone}

\theoremstyle{thmstyletwo}

\theoremstyle{thmstylethree}

\begin{document}

\title[Article Title]{Gendered Cultural Discourse in Japan across the Prewar–Postwar Transition: Evidence from Historical Word Embeddings}

\author[1]{\fnm{Shintaro} \sur{Sakai}}\email{shinsaka@iu.edu}
\author[1]{\fnm{Haewoon} \sur{Kwak}}\email{hwkwak@iu.edu}
\author[1]{\fnm{Jisun} \sur{An}}\email{jisunan@iu.edu}
\author*[2]{\fnm{Akira} \sur{Matsui}}\email{amatsui@rieb.kobe-u.ac.jp}

\affil[1]{\orgname{Indiana University Bloomington}, \orgaddress{\city{Bloomington}, \state{Indiana}, \country{USA}}}

\affil[2]{\orgname{Kobe University}, \orgaddress{\city{Kobe}, \country{Japan}}}

\abstract{
We quantify the evolution of gender stereotypes in Japan from 1900 to 1998, covering the prewar–postwar transition, using a series of yearly word embeddings trained on historical text corpora.
We define the gender stereotype value to measure the strength of a word’s gender association by computing the difference in cosine similarity of the word to female- versus male-related attribute words.
We examine trajectories of gender stereotype across three traditionally gendered domains: Home, Work, and Politics.
To provide a more granular analysis and strengthen the robustness of our findings in the Work domain, we also examine changes in gender stereotypes across 18 occupations and calculate their correlations with gender participation statistics.
Our results reveal domain-specific patterns. In the Home domain, female stereotype values remain stable over time, showing no statistically significant changes. In contrast, the Work and Politics domains exhibit significant trend reversals in female stereotype values around 1945, shifting from negative prewar trends to positive postwar trends, indicating increasing female associations in public domains over time. These findings show that trends in gendered cultural discourse shifted unevenly around 1945: the observed changes in the Work and Politics domains are temporally aligned with postwar institutional transformations, including reforms under the U.S.-led Allied Occupation, while the Home domain shows more limited change. Furthermore, female stereotype values for occupations show a moderately positive correlation (r=0.364) with the proportion of women in each occupation, indicating that word-embedding-based measures of gender stereotype mirrored demographic shifts to a meaningful extent.}


\maketitle

\section{Introduction}\label{intro}
Stereotypes play a central role in maintaining and reproducing social and professional inequalities by shaping how individuals perceive and evaluate others. For example, due to gender stereotypes, identical performances or qualifications are often evaluated differently depending on the perceived gender of the individual \citep{kessler2019incentivized, moss2012science, bowles2007social}. This contributes to persistent gender gaps in occupational outcomes and leadership representation \citep{buffington2016stem, amanatullah2010negotiating, schumann2010women}. Understanding stereotypes helps explain why such disparities persist in society. Because stereotypes shape people’s perceptions and behaviors in ways that sustain inequality, it is also crucial to understand how they evolve over time. Identifying the factors that drive or hinder changes in stereotypes can clarify the mechanisms through which biased beliefs can be reshaped. Insights from such research can inform interventions aimed at promoting more equitable perceptions of social groups and, ultimately, contribute to reducing persistent social and gender disparities.

Traditionally, stereotypes have been measured using methods such as opinion surveys~\citep{katz1933racial, gilbert1951stereotype} and the Implicit Association Test (IAT)~\citep{greenwald1998measuring}. However, these methods rely on human participants, making it impossible to quantify stereotypes from a century ago. While studying changes in stereotypes is also possible by collecting the results of opinion surveys or IAT at different points in time, such data are rarely available across a long period of time or across diverse countries and cultures. Moreover, the reliability of these methods can be compromised by shifts in survey design, sampling biases, and respondents’ reluctance to endorse socially undesirable views~\citep{bergsieker2012stereotyping, bhatia2021changes}.  
To address these limitations, word embeddings have been increasingly used to quantify stereotypes~\citep{garg2018word, kozlowski2019geometry, jones2020stereotypical, bhatia2021changes}. Word embeddings capture stereotypes through distributional patterns of word associations as reflected in written language. By leveraging textual data produced during specific historical periods, they can quantify historical stereotypes even in the absence of survey or psychological test data. Moreover, previous literature has shown that shifts in stereotypes quantified through word embeddings strongly correlate with those measured in traditional social sceince research~\citep{garg2018word, kozlowski2019geometry, jones2020stereotypical, bhatia2021changes}.

While numerous studies have sought to quantify social stereotypes using word embeddings~\citep{caliskan2017semantics, bhatia2017semantic, garg2018word, lewis2020gender, charlesworth2021gender, jones2020stereotypical, kozlowski2019geometry}, the majority focus on Western contexts and largely neglect Asian or non-Western settings.
Among these underexplored non-Western contexts, Japan offers unique analytical value for studying stereotype change because it underwent profound and rapid societal transformations during the 20th century. Following World War I\hspace{-1.2pt}I and the subsequent U.S.-led Allied occupation, Japan experienced a dramatic increase in Western influence, which could have significantly impacted social norms in Japan. Gender roles in the workplace and politics were key areas of these reforms; the U.S.-led Allied occupation (1945--1952) imposed structural changes to dismantle Japan’s patriarchal system to raise women’s status in the workplace and politics~\citep{pharr1981political, molony2018feminism, okuyama2021empowering, van2022equality}. Despite the abrupt introduction of Western-led legal and institutional reforms aimed at restructuring gender hierarchies, there has been no quantitative study showing that these formal changes produced corresponding transformations in widely held gender stereotypes, making it impossible to directly observe how they evolved.

To address this gap, the present study quantifies changes in gender stereotypes from 1900 to 1998, covering much of the 20th century, including the period before and after the U.S.-led institutional reforms implemented in Japan immediately following World War I\hspace{-1.2pt}I.
We focus on this period for two main reasons: (1) its coverage of major historical developments related to women’s participation in society, and (2) the availability of high-quality embeddings derived from corpora publicly released by Japanese government agencies.
We measure gender stereotype changes associated with terms representing the domains of Home, Work, and Politics. Work and Politics were directly targeted by postwar institutional reforms; if these reforms were effective, female associations in these domains would be expected to increase. Home, by contrast, has been traditionally and strongly associated with women, such that a genuine improvement in gender equality would be expected to manifest as a decline in the female-home association. 
Furthermore, to assess whether our embedding-based stereotype measure captures meaningful aspects of gendered social reality, we further analyze 18 occupational terms. We compare the stereotype measures for these occupations with the actual representation of women in the corresponding occupations. These 18 occupational terms were selected for their consistent occurrence across all embedding models for the 1900–1998 period and for having at least one corresponding entry in the Japanese national census.

Our contributions are as follows: (1) To the best of our knowledge, this is the first study to quantify gender stereotypes in the domains of Home, Work, and Politics, as well as across occupations in Japan over nearly the entire twentieth century (1900--1998).
We show that gender stereotypes did not evolve uniformly but diverged sharply across domains: while the Home domain remained largely unchanged, the Work and Politics domains experienced pronounced postwar reversals.  These findings show that trends in gendered cultural discourse in Japan shifted unevenly around 1945, with domain-specific changes temporally aligned with the institutional reforms implemented during the US-led Allied Occupation; and 
(2) We make our trained yearly word embedding models publicly accessible~\citep{wordembeddingmodel}. These models can be used for future research on historical change in Japan, such as the quantification of semantic shifts or other forms of social stereotypes.

\section{Background}\label{sec:background}
This section reviews key seminal studies that establish word embeddings as a robust computational method for quantifying shifts in social stereotypes. We then provide historical context on changes in women's status in Japan. Finally, we discuss the theoretical framework underpinning our hypotheses, drawing on the literature on formal and informal institutions and on uneven gender change across public and private domains.

\subsection{Word embeddings as a tool to measure changes in social stereotypes}
Word embeddings have emerged as a powerful computational tool for capturing historical shifts in social stereotypes. A foundational work by~\citet{garg2018word} quantified changes in gender and ethnic stereotypes in the United States from 1910 to 1990. Their analysis revealed that the changes in occupational female bias in word embeddings closely mirrored actual changes in the percentage of women in those occupations, as recorded by the U.S. Census. This was further validated by historical survey results of gender stereotypes. Similarly, they demonstrated that changes in occupational Asian bias in word embeddings correlated with the changing percentage of Asians in those occupations and aligned with ethnic stereotype survey data. These findings reflect broader cultural shifts, such as the women's movement (1960s-1970s) and the growth of the Asian-American population (1960s-1980s) in the U.S.

Further research has expanded on these insights, exploring different facets of stereotype change. \citet{kozlowski2019geometry}, for instance, examined how gender stereotypes associated with specific occupations like ``engineer'', ``journalist'', and ``nurse'' changed from the 1900s to the 1990s. Their findings indicated that ``nurse'' became less female-associated and ``engineer'' less male-associated over time, yet both occupations retained a persistent gendered stereotype. Interestingly, ``journalist'' transitioned from being male-stereotyped until the early 1950s to becoming female-stereotyped thereafter. These findings align with the broader decline in occupational gender segregation in the 20th century.

Beyond specific occupations, studies have also investigated changes across broader domains. \citet{jones2020stereotypical} quantified shifts in gender stereotypes across ``career'', ``family'', ``science'', and ``arts'', observing a consistent decline in overall gender bias. Despite this, some biases persisted: ``career'', and ``science'' terms remained more strongly associated with men, while ``family'' and ``arts'' continued their stronger association with women. These findings were broadly consistent with observed societal trends.

\citet{bhatia2021changes} approached the topic using established psychological scales to measure gender stereotypes through word embeddings. They concluded that gender biases have generally diminished throughout the 20th century, particularly for stereotypically feminine and personality-related traits. This aligns well with existing psychological theories on gender stereotypes.

More comprehensively, \citet{charlesworth2022historical} explored historical representations of various social groups, including racial, gender, age, body type, and socioeconomic groups, across 200 years of Google Books data (1800-1999). Their extensive analysis revealed that while the specific words and traits most associated with these groups changed significantly over time, the average valence (positivity or negativity) of these stereotypes often remained remarkably stable. This suggests a dynamic evolution of stereotype content even as underlying sentiment persists.

In summary, the growing body of research utilizing word embeddings has shown their effectiveness for quantifying the evolution of social stereotypes across historical periods. These studies demonstrate how word embeddings can capture nuanced changes in gender, ethnic, and other social group stereotypes, often aligning with real-world demographic shifts and validated psychological theories. However, the majority of these studies focus on Western, particularly American, contexts, leaving the question of stereotype change in non-Western societies largely unexplored.

\subsection{The changes in women's status in Japan}
\label{sec:womenrole}
Although women's status had been persistently low in Japan, Japan experienced several reforms by the US-led Allied Occupation after World War I\hspace{-1.2pt}I which challenged traditional gender norms~\citep{pharr1981political, molony2018feminism, okuyama2021empowering, van2022equality}. 

Before 1900, public advocacy for women's rights in Japan was heavily suppressed. Direct feminist activism was rare due to government censorship, and early advocates faced arrest or silencing~\citep{molony2018feminism}. After around 1910, Japanese women began actively challenging traditional gender norms and exploring new forms of self-representation. Feminist publications and media sparked debates on women’s independence, education, sexuality, and professional opportunities. Discussions of issues such as abortion, chastity, and women’s roles in modern occupations reflected a growing engagement with questions of personal autonomy and social participation~\citep{molony2018feminism}. In 1919, the New Women’s Association was founded, demanding suffrage, legal reforms, and labor rights~\citep{molony2018feminism}. Transnational women’s organizations like the Young Women’s Christian Association (YWCA) and the Woman’s Christian Temperance Union (WCTU) played active roles in promoting women’s welfare. However, despite their efforts, women were still denied suffrage, and their working conditions saw little improvement through the prewar period~\citep{garon1998molding}.

Following World War I\hspace{-1.2pt}I, women's status in Japanese politics and the workplace saw significant advancements due to reforms implemented under the US-led Allied Occupation~\citep{molony2018feminism}. In 1946, women were granted the right to vote for the first time, with nearly two-thirds of eligible women participating in the election. That same year, female representatives were elected to the Diet. Some also contributed to drafting the postwar Constitution, which enshrined gender equality as a constitutional principle. Women's labor conditions also improved: more women entered the workforce, and in 1947, the Ministry of Labor established the Women’s and Minors’ Bureau to protect female workers. Article 14 of the 1947 Constitution guarantees equality regardless of gender~\citep{constitution}. The Labor Standards Law of 1947 prohibits gender-based disparities in wages, working hours, and conditions~\citep{laborstandard}. Later, the Equal Employment Opportunity Law of 1985 further advanced gender equality by banning discrimination in hiring and promotion, as well as penalizing disadvantageous treatment due to marriage, pregnancy, or childbirth. 

In summary, Japan underwent substantial formal institutional reforms following World War I\hspace{-1.2pt}I that significantly expanded women’s legal rights and participation in the workplace and politics. These reforms were largely concentrated in public domains, creating new legal frameworks for gender equality in employment and political representation.

\subsection{Interplay between formal and informal institutions}

A central insight in institutional theory is the asymmetry between formal and informal institutions~\citep{north1990institutions, north1994economic, williamson2000new, roland2004understanding}. While formal institutions such as legal rules and policies can be altered relatively quickly, informal institutions, including cultural norms, stereotypes, conventions, and shared understandings, are historically embedded and typically evolve more slowly.

Prior literature consistently highlights this distinction. Formal rules may change rapidly through legislation or political reform, whereas informal constraints are rooted in long-standing social practices and collective belief systems that persist across generations~\citep{north1990institutions, north1994economic}. Institutions can also be understood as operating at multiple levels of social analysis: informal institutions such as customs, traditions, norms, and religion constitute deeply embedded structures that evolve over long historical horizons, while formal political and legal frameworks shift within much shorter time spans~\citep{williamson2000new}.

A related distinction characterizes institutions as either ``slow-moving'' or ``fast-moving''~\citep{roland2004understanding}. Culture, including values, beliefs, and social norms, belongs to the former category, evolving incrementally and continuously and often grounded in enduring religious and ideological systems. Political institutions, by contrast, can change abruptly, sometimes nearly overnight. Because slow-moving institutions gradually build structural constraints, they shape and may resist rapid changes in fast-moving institutions. As a result, formal reforms cannot automatically transform underlying cultural norms.

In summary, this body of work indicates that changes in formal institutions do not necessarily produce immediate transformations in informal institutions. Deeply embedded norms including gender stereotypes are structurally resistant to rapid change, even when legal and political reforms are introduced. This asymmetry is particularly relevant for understanding postwar Japan, where externally imposed legal reforms aimed at gender equality may not have automatically translated into shifts in gendered cultural norms, making it an important empirical question whether and to what extent such reforms reshaped gender stereotypes.

\subsection{Uneven gender role change across public and private domains}

A growing body of research suggests that gender change in actual roles does not occur uniformly across all areas of social life. Instead, changes in gender roles tend to be domain-specific, with more substantial transformations in public spheres such as employment and politics, and more limited change in private domains such as the household. Although these studies are primarily based on U.S. data, they provide important theoretical insights into how gender norms evolve unevenly across domains.

\citet{england2010gender} argues that gender change is fundamentally uneven across domains. While women have increasingly entered the labor market and male-dominated occupations, change in the private or domestic sphere has been much more limited. This asymmetry is explained by differences in incentives: women have strong economic and social incentives to move into higher-status domains, whereas men have weaker incentives to take on lower-status domestic roles. As a result, change has been concentrated in public domains, while traditional norms persist more strongly in the private sphere.

Similarly, \citet{goldin2014grand} documents a long-term convergence in gender roles, particularly in labor market participation, occupational attainment, and education. However, this convergence remains incomplete. Persistent gender gaps—especially in earnings and career progression—are linked to structural features of the labor market, such as the premium on long and inflexible work hours, which interact with women’s greater involvement in family responsibilities. 

Consistent with this perspective, \citet{sayer2011she} show that gender change has been asymmetric, with major shifts in women’s labor force participation but much smaller changes in men’s domestic roles. Despite women’s increased participation in paid work, men’s involvement in household labor has increased only modestly. This imbalance reinforces the persistence of traditional norms in the domestic sphere.

These studies highlight that gender change is uneven and domain-dependent. While public domains such as employment have undergone substantial transformation, private domains remain more resistant to change due to enduring cultural norms and weaker incentives for behavioral shifts. This body of work provides a theoretical foundation for understanding why changes in gender stereotypes may emerge more clearly in public contexts than in domestic ones.

\section{Research hypothesis}\label{sec:hypothesis}

Building upon the background reviewed in Section~\ref{sec:background}, we formulate our research hypotheses as follows. Japan experienced substantial institutional reforms following World War II aimed at improving women’s status in the workplace and politics, including the introduction of women’s suffrage, legal guarantees of gender equality, and labor protections. These reforms primarily targeted public domains and created new opportunities for women’s participation in work and political life. However, from an institutional perspective, changes in formal institutions do not necessarily lead to immediate changes in gender stereotypes. As a result, even significant institutional reforms may have only limited and gradual effects on gender stereotypes. However, the degree to which formal reforms penetrate gender stereotypes is likely to vary across domains, for two reasons. First, postwar reforms directly targeted Work and Politics rather than domestic life, so any effect of formal institutions on gender stereotypes should be most apparent in those domains. Second, research on actual gender role change shows that behavioral shifts tend to be domain-specific, with more substantial transformations in public spheres such as employment and politics, and more limited change in private domains such as the household. Since word embeddings have been shown to track actual social patterns, this behavioral asymmetry would be expected to be reflected in stereotype change as well. Accordingly, we test the following hypothesis: \textbf{H1. Gender stereotypes evolve unevenly across domains, with greater changes in female stereotypes in the Work and Politics domains than in the Home domain between 1900 and 1998.}

In addition to addressing the substantive question of domain-specific stereotype change in H1, H2 examines whether gender stereotypes estimated from historical Japanese texts align with women's occupational representation in Japan. As established in Section~\ref{sec:background}, prior studies have validated word embeddings by showing that stereotype measures correlate with demographic patterns observed in census and survey data. Following this approach, we conduct a complementary occupational-level analysis in the Japanese context. Because occupations are specific social roles for which women's representation can be measured using national census data, they provide a concrete empirical benchmark. We therefore test a second hypothesis: \textbf{H2. Changes in female stereotypes for occupations correspond to the actual gender composition of those occupations as measured by national census data.}

\section{Data and method}
\subsection{Corpus}
The Japanese National Diet Library released the NDL Ngram Dataset~\citep{ngramdata} in 2022. The dataset provides book and magazine data, covering publications released in Japan since the Meiji era (around 1868). It includes uni-grams to five-grams along with their yearly frequencies. N-grams that appear fewer than four times through the whole periods were excluded during corpus construction. Although the dataset aims to include all books and magazines published in Japan, it only contains works whose copyright protection has expired. We merged the book and magazine datasets to train yearly embedding models for 1900--1999. We also trained separate models on the book-only and magazine-only corpora. However, as these models were of lower quality than those based on the merged data, we use only the latter for our main analysis. Supplementary analyses using the book-only and magazine-only corpora are provided in the Appendix~\ref{sec-book-analysis} and~\ref{sec-magazine-analysis}.

\begin{figure*}[h]
\centering
\includegraphics[width=1.0\textwidth]{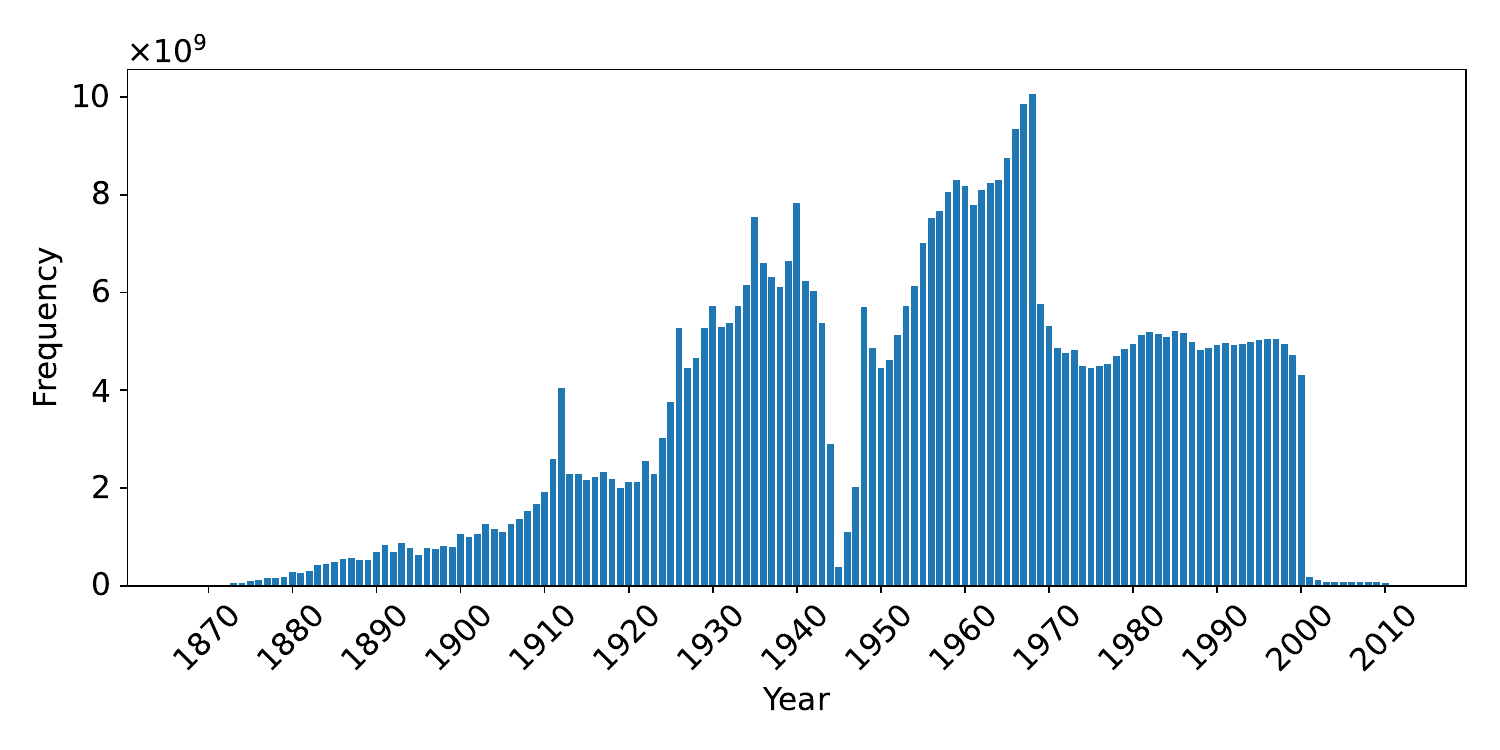}
\caption{\label{fig:frequencyall}The total number of n-grams for each year in the merged NDL Ngram dataset (books and magazines).}
\end{figure*}

Figure~\ref{fig:frequencyall} presents the total number of n-grams per year. 
We restricted our training to the period between 1900 and 1999 for two reasons: (1) consistency with previous studies that quantify stereotypes over a 100-year span \citep{garg2018word, kozlowski2019geometry, bhatia2021changes}, and (2) sufficient temporal coverage both before and after the U.S.-led Allied occupation of Japan.
Notably, the total number of n-grams in 1945 drops sharply, likely reflecting the disruptions caused by World War I\hspace{-1.2pt}I and postwar censorship under the Occupation. The sharp decline in n-grams beginning around 1970 is the result of copyright limitations on which works could be digitized.

Using metadata from the Japanese National Diet Library, which include information on book titles, authors, and genre classifications, we estimated (1) the distribution of book genres and (2) the proportion of male and female authors. We find that no single genre accounts for a majority of the books, and that genre distributions remain highly stable across time. In addition, between 66\% and 92\% of authors are inferred to be male across the years between 1900 and 1999. Together, these results suggest that temporal changes in our estimates are unlikely to be driven solely by large shifts in genre composition. The strong male predominance among authors may also reflect the gendered structure of literary production in Japan during this period, rather than constituting only a source of bias.
Detailed methods and results are provided in the Appendix~\ref{secbookcategory} and~\ref{secauthor}. Note that genre and author name information are not available for the magazine data.

\subsection{Preprocessing and model training}
Before training the word embeddings, we first performed word segmentation on the bigram to five-gram data. Word segmentation is a necessary preprocessing step for Japanese natural language processing (NLP), as Japanese is written without spaces between words. We used MeCab~\citep{kudo2005mecab} in combination with a dictionary designed for historical Japanese texts~\citep{ogiso2013morphological, unidic} for this task. As part of preprocessing, we also removed n-grams composed solely of alphabets, numbers, or postpositional particles, as they carry limited semantic content in Japanese.

We trained a separate word embedding model for each year from 1900 to 1999 using the skip-gram with negative sampling (SGNS) algorithm~\citep{mikolov2013linguistic}. SGNS aims to maximize the probability of observed word–context pairs in the corpus while minimizing the probability of randomly sampled, unobserved pairs (i.e., negative samples). We used SGNS due to its widespread adoption in prior research~\citep{garg2018word, kozlowski2019geometry, jones2020stereotypical, bhatia2021changes, charlesworth2021gender}. The window size was set to 4 because the corpus includes only up to five-gram sequences. We used a negative sampling of 10 to accelerate the training process and set the vector size of 300 in accordance with commonly used pre-trained historical word embeddings~\citep{hamilton2016diachronic}. For implementation, we used the Python wrapper for the original C++ code provided by~\citet{zhao2017ngram2vec}~\citep{ngram2vec}.

\subsection{Quantification of gender stereotype}
In this study, we define gender stereotypes as patterns of gendered word associations in language, that is, the extent to which words representing specific domains (e.g., Home, Work, and Politics) are more strongly associated with female or male attribute terms in text. To calculate gender stereotypes, we relied on a widely used bias measurement method called the Word Embedding Association Test (WEAT) proposed by~\citet{caliskan2017semantics}. The WEAT defines stereotype in a word embedding model as the difference between the mean cosine similarity of two lists of attribute words to a target word in a given domain, as follows:

\begin{align}
\label{eq:1}
S(X) = \frac{1}{|X|} \sum_{x \in X} s(x, A, B)
\end{align}
where
\begin{equation}
\begin{aligned}
s(x, A, B) &= \frac{1}{|A|} \sum_{a \in A} \cos(v_x, v_a) \\
&\quad - \frac{1}{|B|} \sum_{b \in B} \cos(v_x, v_b)
\end{aligned}
\end{equation}

\noindent
$v_{i}$ is a vector of word $i$ and $\text{cos} \left( v_{i},v_{j}\right)$ is a cosine similarity between $v_{i}$ and $v_{j}$. $x$ is a target word in list $X$. $A$ and $B$ are lists of attribute words. 

In our study, target words are the words representing the Home, Work, or Politics domain, or occupations, and two lists of attribute words $A$ and $B$ are Japanese gender words for women and men, respectively. For example, list $A$ includes \Ja{女}(woman) and \Ja{女性}(female), while list $B$ includes \Ja{男}(man) and \Ja{男性}(male). A positive WEAT score indicates that the target word or the domain is female-coded (more strongly associated with female attribute terms), whereas a negative score indicates that it is male-coded (more strongly associated with male attribute terms). Accordingly, a change in the score indicates a directional shift in gender association --- toward more female-coded or more male-coded --- rather than an absolute increase or decrease in stereotype strength.

Following the work by~\citet{garg2018word}, we address the known sensitivity of WEAT-style measures to word list composition by employing a bootstrap procedure when calculating stereotype values for each year within the Home, Work, and Politics domains. Specifically, in each iteration, domain and gender words are randomly resampled with replacement to match the original domain size, and the stereotype value is recalculated using Equation (\ref{eq:1}).
This process is crucial because it quantifies the uncertainty of stereotype value estimates within each domain, especially when domain sizes are small and individual words have a large influence. This entire process is repeated 1,000 times for each year, producing a distribution of scores that reflects both the uncertainty and stability of the domain-specific results. From this distribution, we derive 95\% confidence intervals around our estimates.

\subsection{Construction of word lists for quantifying gender stereotypes}
To collect words representing Home and Work domains, we drew on the J-LIWC2015 dictionary~\citep{igarashi2022development}, the latest Japanese version of the LIWC dictionary which is based on LIWC2015 (Linguistic Inquiry and Word Count 2015) dictionary~\citep{pennebaker2015development}. The LIWC dictionary is one of the most widely used and validated dictionaries, undergoing multiple reliability and validity assessments~\citep{tausczik2010psychological}. 
The LIWC dictionary categorizes words into various categories such as linguistic dimension and psychological construct. 

The J-LIWC2015 dictionary follows the structure of the original LIWC2015 and was carefully developed through qualitative evaluations by psychologists, internal consistency checks of the categories, and human validation. From this dictionary, we collected words for the Home and Work domains from ``home'' and ``work'' categories, which contains 148 and 837 words, respectively. Because J-LIWC2015 does not include a Political category, we instead used Empath, an open-source and validated dictionary~\citep{fast2016empath}. We first extracted all the English words in the politics category and manually translated them into Japanese. 

We removed words from each domain that were not present in all 100 word embedding models to ensure comparability across time. Importantly, we also manually excluded explicitly gendered terms within the Home domain, such as \Ja{主婦} (housewife) and \Ja{主人} (husband), because these terms lexically encode gender by definition. Including such words would conflate implicit domain-level gender associations with overtly gendered role labels. Since our objective is to examine how theoretically gender-neutral domains become implicitly gendered in language, we restrict our analysis to terms that do not themselves explicitly specify gender. To verify that this choice does not materially affect the results, we conducted a sensitivity check comparing the original Home word list with an extended version that includes these explicitly gendered terms (\Ja{主婦} (housewife), \Ja{乳母} (wet nurse), \Ja{主人} (husband), \Ja{亭主} (husband), \Ja{女中} (maid)); the two specifications produce nearly identical trajectories throughout the entire 1900--1998 period (Figure~\ref{fig:sensitivity_home}). We did not find any gendered words for Work and Politics domains. The final list includes 85 words for Home, 551 words for Work, and 35 words for Politics domains. Table~\ref{table:genderwords} shows the lists of example words in each domain.

We also used the J-LIWC2015 dictionary to collect gender word pairs by referring to ``Female references'' and ``Male references'' categories. As with the selection of words for the Home and Work domains, words that do not exist in all of the 100 word embedding models were excluded from the lists. The final lists contain 20 gender word pairs. Table~\ref{table:genderwords} shows the lists of example gender words.

Since J-LIWC is a copyright-protected library, we show only a few example words in Table~\ref{table:genderwords}. To ensure reproducibility, we provide the corresponding row numbers of the words in the J-LIWC dictionary file in Table~\ref{table:LIWCrownumber}, and the complete list of words in the Politics domain in Table~\ref{table:politicswords} in the Appendix~\ref{sec-wordlist}.

\begin{table*}[h]
\centering\small
\caption{The lists of example male, female, home, work, and politics words.}
\label{table:genderwords}
\begin{tabularx}{\textwidth}{X X X X X}
\toprule
 Home words & Work words & Politics words & Male words & Female words \\
\midrule
\Ja{家} (home/house), \Ja{台所} (kitchen), \Ja{園芸} (gardening), \Ja{掃除} (cleaning), \Ja{近所} (neighborhood) &
\Ja{仕事} (work), \Ja{マーケット} (market), \Ja{顧客} (client), \Ja{上司} (boss), \Ja{交渉} (negotiation) &
\Ja{国家} (nation), \Ja{知事} (governor), \Ja{政治家} (politician), \Ja{選挙} (election), \Ja{活動家} (activist) &
\Ja{彼} (he), \Ja{男} (man), \Ja{男性} (male), \Ja{男児} (boy), \Ja{男の子} (boy) &
\Ja{彼女} (she), \Ja{女} (woman), \Ja{女性} (female), \Ja{女児} (girl), \Ja{女の子} (girl) \\
\bottomrule
\end{tabularx}
\end{table*}

For occupations, we first collected occupations from the Japan Standard Occupational Classification (1960)~\citep{occuclass}. To ensure robustness for our analysis, we applied two selection criteria to filter out some occupations from the original list. First, occupations had to consistently appear across all 100 embedding models. Second, they needed to have at least one data point in the Japanese national census~\citep{nationalcensus}. Since the census is conducted every five years and gender-disaggregated statistics are only available after 1940, with the statistics for 1985, 1990, and 1995 missing, the maximum number of data points available for any occupation was nine. Table~\ref{table:occupationalwords} lists the 18 occupations that satisfied both criteria.

\begin{table*}[h]
\centering\small
\caption{The lists of occupational words.}
\label{table:occupationalwords}
\begin{tabularx}{\textwidth}{X}
\toprule
{} Occupational words \\
\midrule
\Ja{医師} (doctor), \Ja{歯科医師} (dentist), \Ja{画家} (painter), \Ja{音楽家} (musician),
\Ja{俳優} (actor), \Ja{弁護士} (lawyer), \Ja{記者} (journalist),  \Ja{駅長} (stationmaster),  \Ja{船長} (ship captain), \Ja{車掌} (train conductor), \Ja{船頭} (boatman), \Ja{木工} (woodworker), \Ja{石工} (stonemason),  \Ja{大工} (carpenter), \Ja{左官} (plasterer), \Ja{土工} (construction worker),  \Ja{芸者} (Geisha), \Ja{料理人} (cook) \\
\bottomrule
\end{tabularx}
\end{table*} 

\subsection{Interrupted time-series analysis (ITS)}

To examine whether the trend in female stereotypes changed around 1945, we applied an interrupted time-series analysis (ITS).
Interrupted time-series analysis provides a useful framework for detecting abrupt changes while accounting for underlying time trends. By comparing observations just before and after 1945, this method allows us to assess whether the period following the US-led Allied Occupation corresponded to a statistically significant shift in gender stereotypes, rather than being part of a continuous long-term trend.
In this study, the dependent variable $S_t$ represents the female stereotype value (directional gender association score) in each year $t$, measured between 1900 and 1999, and the cutoff year $t = 1945$ corresponds to the end of World War I\hspace{-1.2pt}I, a historically significant turning point in Japanese society. The model takes the following linear form:  

\begin{equation}
S_t = \alpha_0 + \beta_0 t + \alpha_i D_t + \beta_i D_t t + \epsilon_t
\end{equation}

In this model, $D_t$ is a binary indicator that takes the value 1 for years after 1945 and 0 for years before 1945. This formulation enables us to examine two distinct aspects of change around the cutoff year: the immediate level shift, represented by the coefficient $\alpha_i$, and the alteration in the temporal trend, captured by the coefficient $\beta_i$. Our primary interest lies in $\beta_i$, which indicates whether the rate or direction of change in female stereotypes differs significantly between the pre- and post-1945 periods. The term $\epsilon_t$ denotes the random error component at time $t$. By applying this model to the time series of female stereotype values across different domains (Home, Work, and Politics), we can test whether postwar societal transformations corresponded to statistically significant shifts in gender stereotypes.

\section{Preliminary analysis}\label{sec:preliminary}
Prior to the historical analysis of gender stereotypes, we first ensured the quality of the word embedding models trained for each year. For this, we used JWSAN, a large-scale Japanese dataset designed to evaluate distributional semantic models in terms of word similarity and word association \cite{inohara2022jwsan}. Word similarity captures semantic closeness in meaning (e.g., synonyms), whereas word association reflects conceptual or functional relatedness. For each word pair in the JWSAN-1400 dataset, we calculated cosine similarity using the embedding models and compared these values with human-annotated similarity and association scores in the dataset by computing Pearson correlation coefficients.

Overall, the models demonstrate consistently stable quality between 1900 and 1998, showing no significant deterioration. On average, the similarity score is 0.416 and the association score is 0.500 (Figures~\ref{fig:quality_sim_combined} and~\ref{fig:quality_assoc_combined}). We also find that yearly similarity and association scores are strongly correlated with the data size of the corpus (Figures~\ref{fig:quality_sim_corr_combined} and~\ref{fig:quality_assoc_corr_combined}), suggesting that data size had a substantial impact on the quality of the yearly word embedding models.

Importantly, the quality scores show a marked decline in 1999. Therefore, to ensure our conclusions are drawn from reliable word embeddings, we restrict our core analysis and statistical tests to the period between 1900 and 1998, where model quality is stable. 
While our charts display the full temporal trend for broader context, the year 1999 is shaded. Interpretation of this region requires caution, as it is derived from a lower-quality embedding model. Values for 1999 are shown for contextual purposes only and are excluded from subsequent analyses.

\section{Results}\label{sec:result}
This section presents the main findings from our analysis, organized around the two hypotheses introduced in Section~\ref{sec:hypothesis}. H1 examines how female stereotypes in the Home, Work, and Politics domains evolved between 1900 and 1998, with particular attention to changes before and after World War I\hspace{-1.2pt}I. H2 investigates whether changes in female stereotypes across 18 occupations correspond to actual changes in women’s labor force participation as reflected in national census data.

\subsection{H1. Gender stereotypes evolve unevenly across domains, with greater changes in female stereotypes in the Work and Politics domains than in the Home domain between 1900 and 1998.}
\begin{figure}[htbp]
\includegraphics[width=1.0\textwidth]{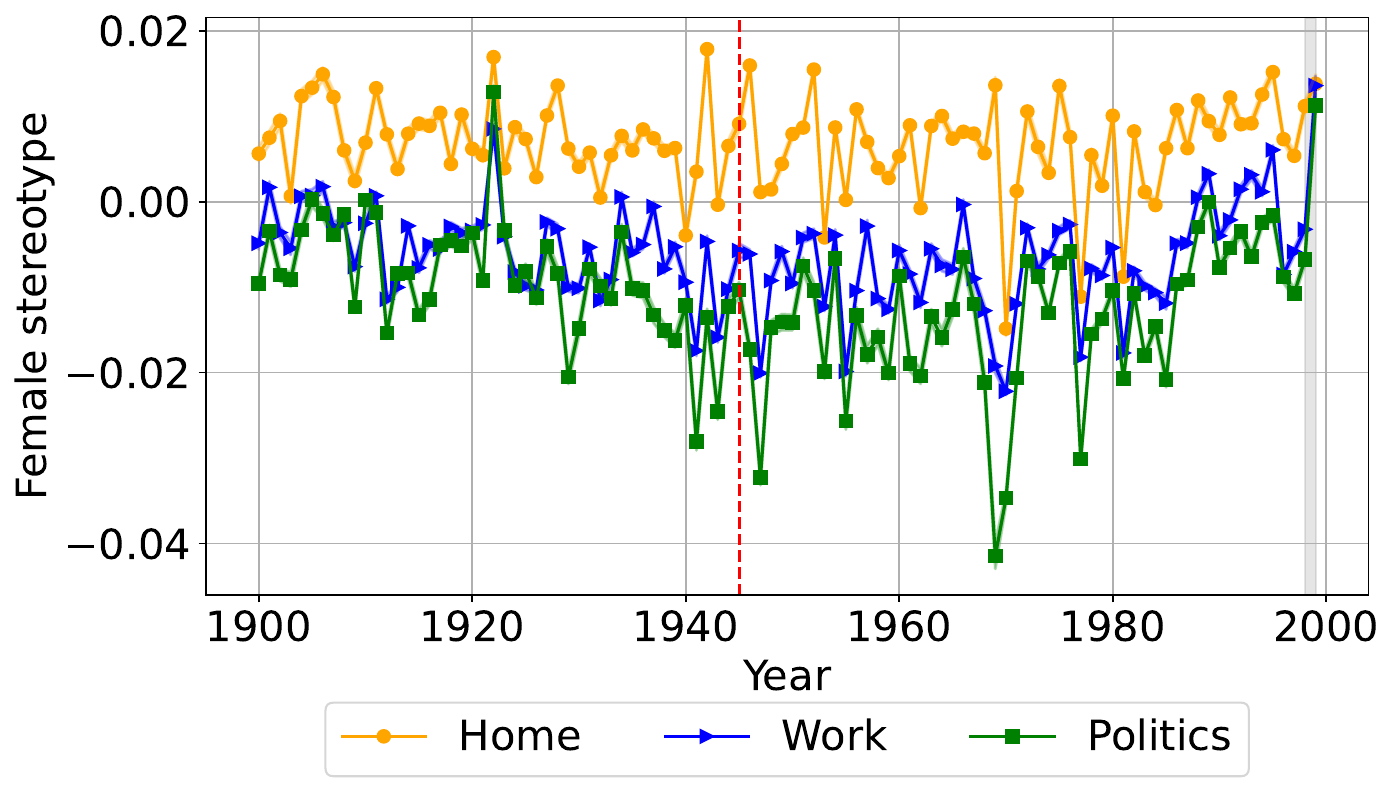}
\caption{\label{fig:workfamilypolitics}The changes of female stereotypes of Home, Work, and Politics over the century. The red vertical line corresponds to 1945, the year World War I\hspace{-1.2pt}I ended. We also include shaded regions representing the 95\% confidence intervals around the stereotype estimates for each domain. However, the intervals are not visible in the plots because their width is negligible compared to the magnitude of the overall stereotype change. The shaded areas represent the periods outside the range between 1900 and 1998.}
\end{figure}

Figure~\ref{fig:workfamilypolitics} shows the changes in female stereotypes in the Home, Work, and Politics domains. 
The shaded area in the figure represents the year 1999, as we explained in Section~\ref{sec:preliminary}.
To quantify overall changes before and after 1945, following the approach of~\citet{jones2020stereotypical} and~\citet{charlesworth2022historical}, we estimated the trend coefficient for each domain as the per-year change in female stereotypes using linear regression for the periods 1900–1944 and 1946–1998.

\begin{figure}[ht]
\centering
\begin{minipage}{0.32\textwidth}
    \centering
    \includegraphics[width=\linewidth]{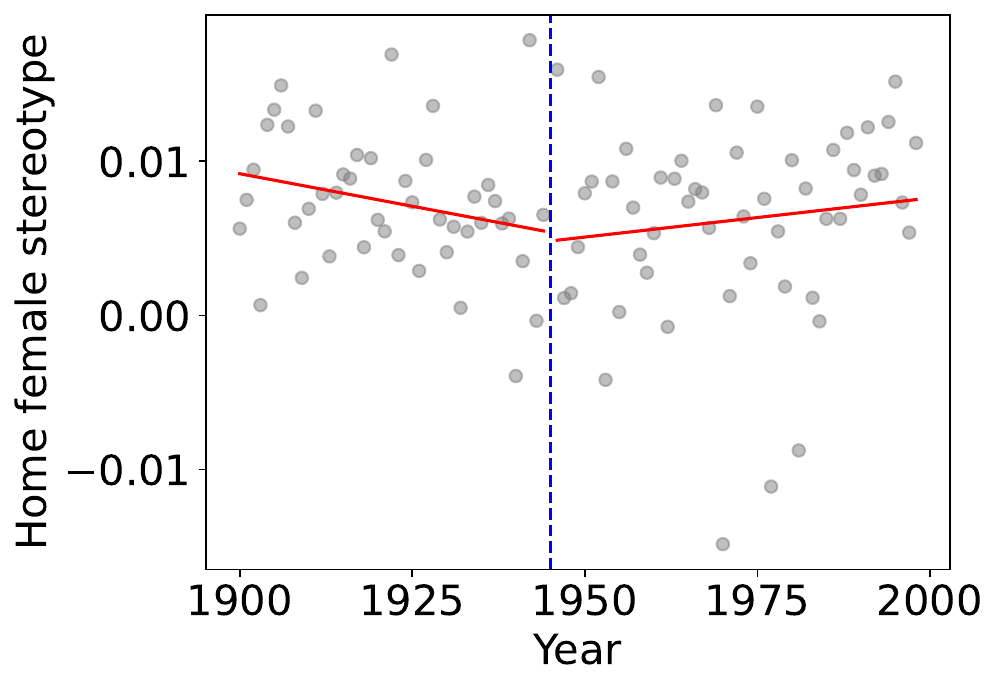}
    \caption*{(a) Home}
\end{minipage}
\hfill
\begin{minipage}{0.32\textwidth}
    \centering
    \includegraphics[width=\linewidth]{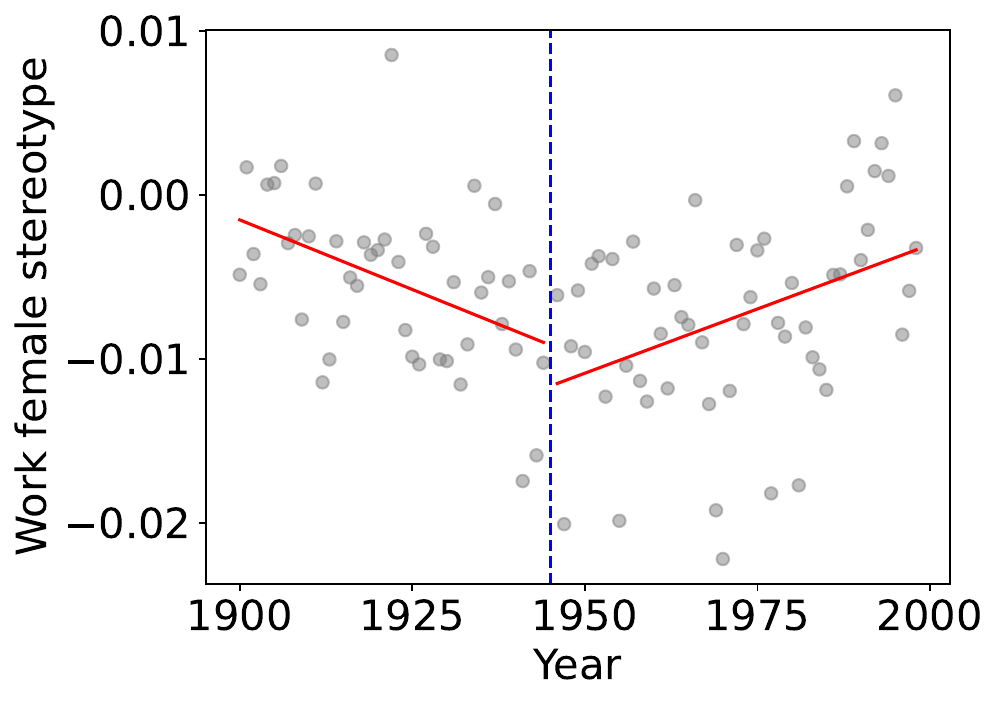}
    \caption*{(b) Work}
\end{minipage}
\hfill
\begin{minipage}{0.32\textwidth}
    \centering
    \includegraphics[width=\linewidth]{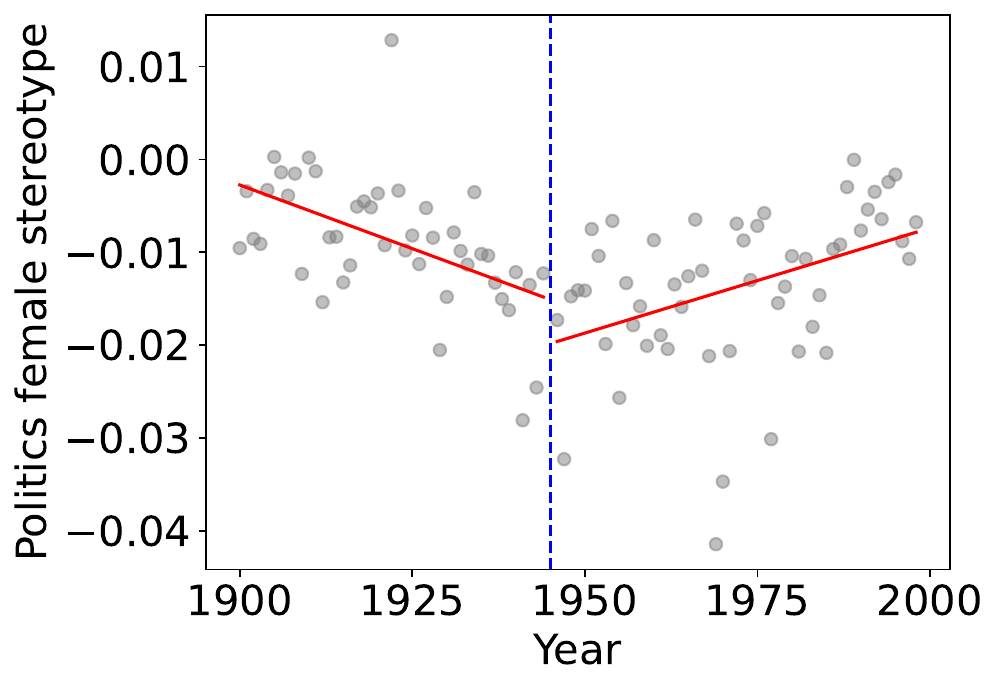}
    \caption*{(c) Politics}
\end{minipage}
\caption{Trends in female stereotypes for the Home, Work, and Politics domains across two historical periods (1900--1944 and 1946--1998). The red lines represent a fit line for each period and the vertical dashed line represents the 1945 cutoff.}
\label{fig:trend_rdd}
\end{figure}

The temporal trends in female stereotypes before and after 1945 reveal divergent patterns across the three domains (Table~\ref{tab:trend_split}). In the Home domain, changes in trends are not statistically significant in either period, indicating relative stability over time. In contrast, the Work and Politics domains exhibit statistically significant trends in both periods, shifting from negative prewar trends to positive postwar trends (all $p < 0.01$). The interrupted time-series (ITS) analysis further supports this pattern (Table~\ref{tab:rdd_trend_change} and Figures~\ref{fig:trend_rdd}). While the Home domain does not exhibit a statistically significant change in trend at the 1945 cutoff ($p = 0.092$), both the Work and Politics domains show significant positive changes in trend (both $p < 0.001$), indicating that the postwar trajectories became significantly more positive. These results are temporally aligned with the institutional transformations following 1945 and are consistent with the historical argument that such transformations corresponded to changes in how women are represented in language in work and political contexts, while the Home domain remained relatively stable.

Furthermore, the overall level of female stereotypes remains clearly distinct across domains. The Home domain exhibits the strongest and most persistent female associations throughout the entire period. In contrast, the Work and Politics domains display not only lower but largely negative female stereotype values throughout most of the period, indicating that these domains have been predominantly male-stereotyped in language. This is consistent with the historical reality that the workplace and politics were male-dominated spheres for most of the twentieth century~\citep{yamaguchi2016determinants, yamaguchi2019impediments, eto2020women}. Although the postwar trend reversals indicate a gradual movement toward less male-stereotyped associations in Work and Politics, these domains remained overall male-coded. For example,  in corporate settings, human resource systems continue to favor men for promotions, resulting in only 9.1\% of women in managerial positions compared to 42.5\% in the United States in 2005, alongside a substantial gender wage gap~\citep{yamaguchi2019impediments, yamaguchi2016determinants}. In politics, women held only 10\% of seats in the national Lower House — a figure that has barely changed since women gained the right to vote in 1946~\citep{eto2020women}. Additonally, the enduringly high female stereotype levels in the Home domain reflect the long-standing cultural association of women with domestic roles~\citep{yamaguchi2016determinants, yamaguchi2019impediments}.

\begin{table}[ht]
\centering
\caption{Estimated per-year change in female stereotypes for Home, Work, and Politics domains across two historical periods (1900--1944 and 1946--1998). $^{***}p<0.001$, $^{**}p<0.01$, $^{*}p<0.05$.}
\label{tab:trend_split}
\begin{tabularx}{\textwidth}{l *{4}{>{\centering\arraybackslash}X}}
\hline
\textbf{Domain} & \textbf{Trend Coefficient (1900--1944)} & \textbf{P-value (1900--1944)} & \textbf{Trend Coefficient (1946--1998)} & \textbf{P-value (1946--1998)} \\
\hline
Home
& $-8.44 \times 10^{-5}$
& 0.097
& $5.07 \times 10^{-5}$
& 0.370 \\

Work
& $-1.70 \times 10^{-4}$
& 0.002**
& $1.57 \times 10^{-4}$
& 0.004** \\

Politics
& $-2.74 \times 10^{-4}$
& 0.0002***
& $2.26 \times 10^{-4}$
& 0.002** \\
\hline
\end{tabularx}
\end{table}

\begin{table}[ht]
\centering
\caption{Estimated trend changes in female stereotypes for Home, Work, and Politics domains based on the interrupted time-series analysis (ITS) with a 1945 cutoff. Trend change coefficients ($\beta_i$) represent the difference in trend coefficient before and after 1945. $^{***}p<0.001$, $^{**}p<0.01$, $^{*}p<0.05$.}
\label{tab:rdd_trend_change}
\begin{tabularx}{\textwidth}{l *{3}{>{\centering\arraybackslash}X}}
\hline
\textbf{Domain} & \textbf{Trend Change Coefficient ($\beta_i$)} & \textbf{P-value} \\
\hline
Home
& $1.35 \times 10^{-4}$
& 0.092 \\

Work
& $3.26 \times 10^{-4}$
& 0.000035*** \\

Politics
& $5.00 \times 10^{-4}$
& 0.000004*** \\
\hline
\end{tabularx}
\end{table}

\subsection{H2. Changes in female stereotypes for occupations correspond to actual gender composition in those occupations as measured by national census data.}

\begin{figure}[htbp]
\includegraphics[width=1.0\textwidth]{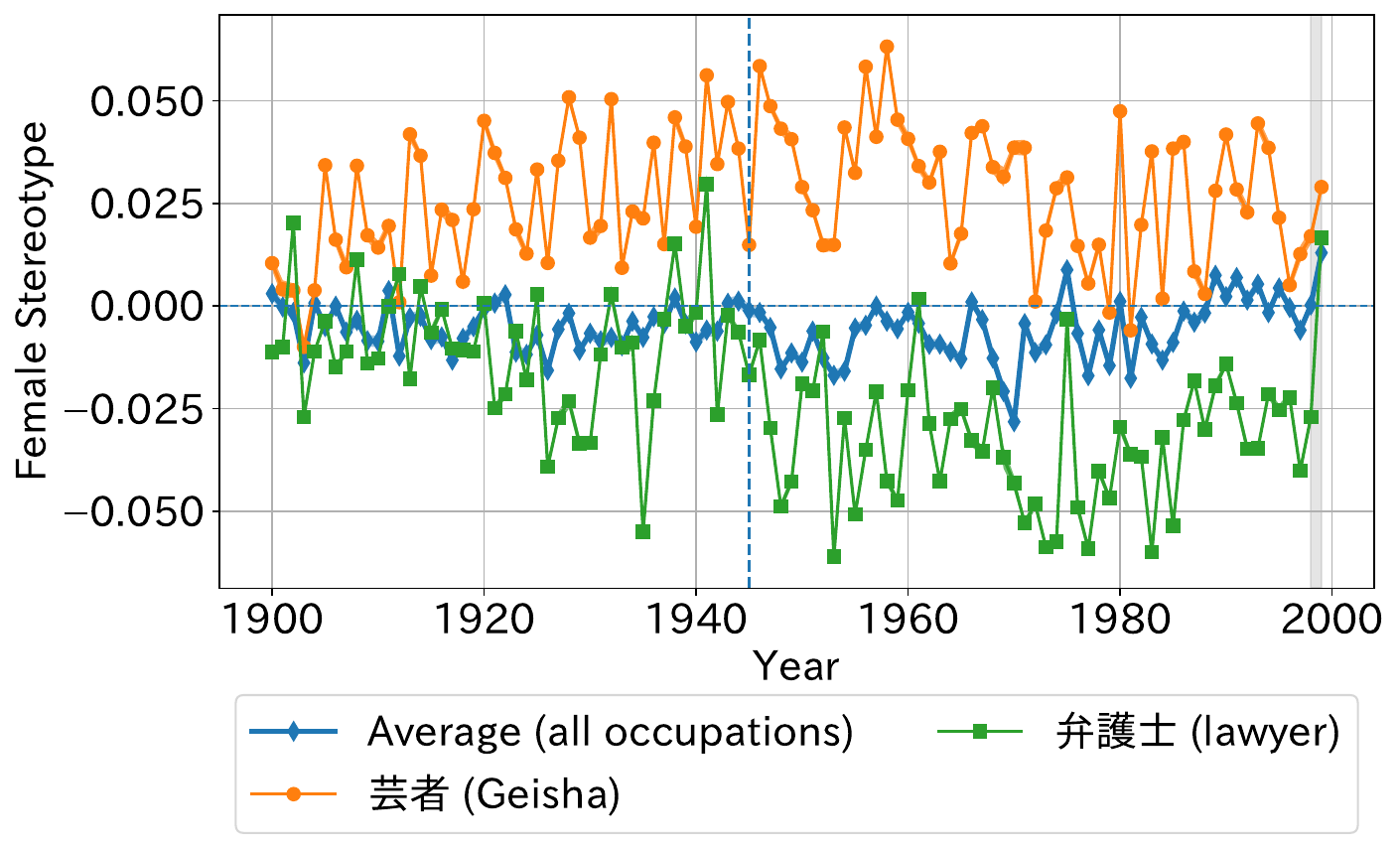}
\caption{The change in female stereotypes averaged across all 18 occupations over the century, along with the female stereotype changes for \Ja{弁護士} (lawyer) and \Ja{芸者} (Geisha). The red vertical line in the figure corresponds to 1945, the year World War I\hspace{-1.2pt}I ended. The shaded areas represent the periods outside the range between 1900 and 1998.}\label{fig:occupations_highlighted}
\end{figure}

Figure~\ref{fig:occupations_highlighted} illustrates the change in female stereotypes averaged across all 18 occupations over the century, along with the female stereotype changes for \Ja{弁護士} (lawyer) and \Ja{芸者} (Geisha), which exhibit the lowest and highest average levels of female stereotype across the period, respectively. To maintain clarity, we display only the average and these two occupations. For the stereotype trajectories of all individual occupations, please refer to Section~\ref{secA3} in the Appendix.

Overall, the average female stereotype value for the 18 occupations was mostly negative between 1900 and 1998, indicating an overall prevailing male stereotype for occupations in general over the century. We highlight \Ja{弁護士} (lawyer) and \Ja{芸者} (Geisha) as representative cases. These patterns are consistent with historical and social context. Lawyer is a historically male-dominated occupation. For example, in 1970, women accounted for only 2.4\% of lawyers, judges, and prosecutors. In contrast, Geisha refers to a traditional Japanese cultural profession historically performed almost exclusively by women. Geisha are female entertainers trained in classical arts such as music, dance, and conversation, and the occupation is institutionally and socially defined as female. The strong female association observed for Geisha therefore aligns with its gender-exclusive nature, while the strong male association for lawyer reflects its male dominance.

Following the analysis for H1, we estimated the per-year trend coefficients for the average of 18 occupations. Between 1900 and 1944, the trend coefficient was slightly negative ($-2.34 \times 10^{-5}$) and not statistically significant ($p = 0.681$), indicating that female stereotypes in these occupations remained largely stable during the prewar period. In contrast, between 1946 and 1998, the estimated trend coefficient was positive ($1.98 \times 10^{-4}$) and statistically significant ($p = 0.003$), suggesting a meaningful upward shift in female stereotype values during the postwar period. This pattern is temporally aligned with broader institutional and societal transformations after 1945.

\begin{figure}[htbp]
\includegraphics[width=1.0\textwidth]{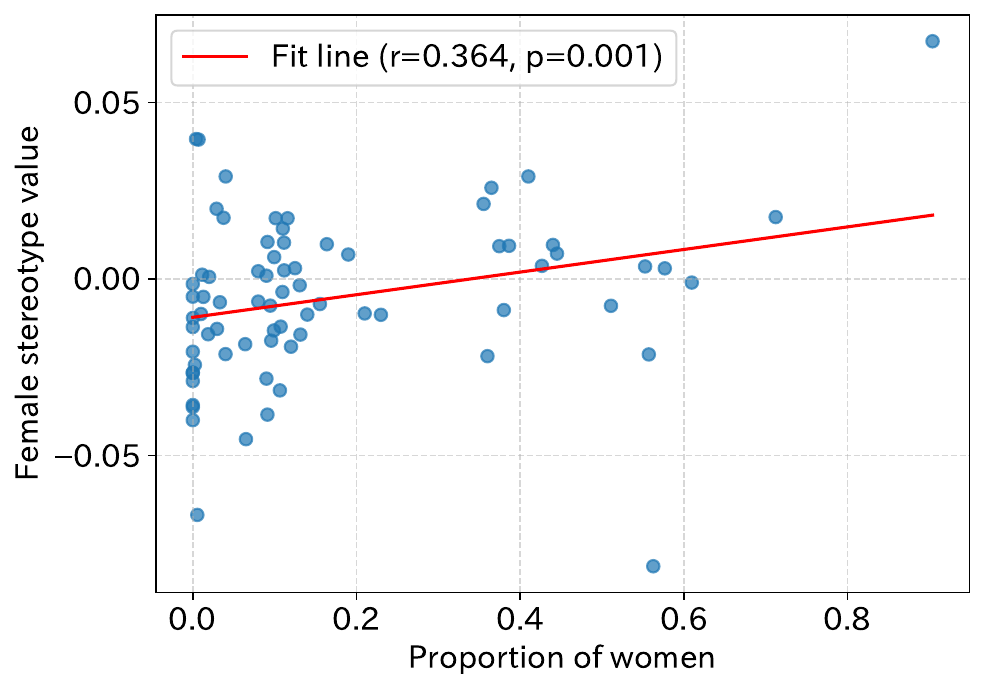}
\caption{\label{fig:corr_occupations_all}The overall correlation between the proportion of women in each occupation-year and the corresponding female stereotype value.}
\end{figure}

To examine the relationship between occupational gender representation and gender stereotypes derived from word embeddings, we conducted a correlation analysis between the proportion of women and the corresponding female stereotype values for the years between 1900 and 1998. First, we calculated the overall correlation by pooling all occupations and years, correlating the proportion of women in each occupation-year with its female stereotype value, and computing the Pearson correlation coefficient ($r$) along with its statistical significance ($p$) (Figure~\ref{fig:corr_occupations_all}). This provided a general measure of the association between women’s representation and stereotype values across the entire dataset. Across all occupations and years, we observed a moderate positive correlation between the proportion of women and the female stereotype value, which was statistically significant ($r = 0.364$, $p = 0.001$). These results suggest that, overall, higher female representation within an occupation is associated with stronger alignment with female stereotypes. While we performed the same analysis separately for each occupation, all the occupations did not reach statistical significance due to insufficient data points in the national census.

\section{Discussion and conclusion}\label{discussion}
The overall patterns reveal domain-specific dynamics in the change of gender stereotypes, supporting H1. The Work and Politics domains exhibit clear trend reversals around 1945, shifting from negative prewar trends to positive postwar trends in female stereotype. Since Work and Politics are predominantly male-coded domains throughout the period, this postwar shift reflects a domain-specific movement away from male-coded associations in public domains. Such shifts are consistent with broader institutional and societal transformations introduced under the U.S.-led Allied occupation, including women’s suffrage, labor protections, and constitutional guarantees of equality, which expanded women’s participation in public life.

The prewar decline in female association in the Work and Politics domains also warrants explicit attention. Between 1900 and 1944, female stereotype values in these domains showed a statistically significant negative trend, meaning that Work and Politics became increasingly male-coded in the decades leading up to 1945. We consider two explanations for this prewar trend, each consistent with the observed pattern, though the present analysis cannot distinguish their relative contributions. First, wartime censorship and political surveillance increasingly constrained autonomous feminist activism from the late 1930s onward. As Japan mobilized for war, many women’s organizations were either suppressed or incorporated into state-sponsored mobilization efforts, shifting public representations of women away from political agency and toward patriotic service and domestic responsibility~\citep{mackie1997creating}, and further reducing the presence of female-coded expression in the corpus~\citep{shillony1991politics}. Figure~\ref{fig:author-geder-composition} shows a gradual decline in female authorship during the prewar period, which is consistent with this explanation; however, we note that the data alone do not establish that the decline in female authorship was caused by wartime censorship and political surveillance specifically. Second, the intensification of nationalist and militarist discourse from the early 1930s likely altered corpus composition in ways that mechanically reduced measured female association in both domains. The Manchurian Incident of 1931 marked a decisive shift in Japan's political climate. Nationalism replaced liberalism as the predominant ideology, and the military became the dominant group in the state~\citep{shillony1991politics}. Subsequently, the National Mobilization Law of 1938 empowered the government to suppress undesired publications while expanding the output of military and nationalist content~\citep{shillony1991politics}. Several terms in the Politics and Work domain word lists --- including \Ja{勅令} (imperial decree), \Ja{国家} (nation-state), \Ja{議員} (Diet member), and \Ja{製造業} (manufacturing industry) --- were institutionally associated with male authority throughout the prewar period; an expansion of publications centered on these terms would mechanically reduce measured female association. Disentangling these mechanisms would require evidence beyond what word embeddings can provide.

In contrast, the Home domain shows no statistically significant change in trends, indicating that female stereotype values remained stable throughout the period. More precisely, female associations expanded across public domains while domestic associations remained stable. One interpretation of this pattern is that women's representations broadened across domains rather than shifted from one role to another.

Importantly, although Work and Politics show significant trend reversals, these changes do not lead to a complete convergence across domains. Female stereotype levels remain consistently highest in the Home domain throughout the entire period, whereas Work and Politics exhibit lower and more comparable levels. Such findings are consistent with prior research showing that women in Japan remain underrepresented in both the workplace and politics~\citep{yamaguchi2016determinants, yamaguchi2019causes, yamaguchi2019impediments, eto2020women}, highlighting the continued influence of structural and cultural constraints.

We note, however, that the present analysis cannot distinguish between three related mechanisms: the observed changes may reflect the direct effect of institutional reforms, demographic shifts in women's actual participation in work and political life, or broader transformations in cultural discourse. Moreover, these mechanisms are not mutually exclusive and likely interact with one another, and our analysis cannot causally identify which mechanism, or combination of mechanisms, drives the observed patterns. All three are consistent with the observed temporal patterns, and disentangling them would require data beyond what historical word embeddings can provide.

Our analysis of occupations provides several key insights. First, when averaging across all 18 occupations, we find that occupational gender stereotypes remained largely stable during the prewar period (1900–1944), with no statistically significant trend. In contrast, the postwar period (1946–1998) shows a statistically significant positive trend, indicating a gradual increase in female stereotype values over time. While this pattern could partly depend on the specific selection of occupations included in the analysis, it nonetheless suggests that, although occupations were historically male-stereotyped and exhibited little change prior to 1945, they began to shift meaningfully in the postwar period. This pattern is temporally aligned with broader institutional and societal transformations after 1945 and is consistent with a gradual reconfiguration of gender representations in professional domains.

Second, the correlation analysis demonstrates that, when pooling across all occupations and years (1900–1998), occupations with higher female representation were associated with stronger female stereotypes in language. Specifically, we observe a moderate positive and statistically significant correlation between the proportion of women and female stereotype values (Pearson’s $r = 0.364$, $p = 0.001$). This finding suggests that word-embedding-based measures capture meaningful relationships, to some extent, between occupational gender composition and linguistic representations. This finding also speaks to the cross-cultural applicability of embedding-based stereotype measures. Although embedding-based stereotype measures have been tested on English-language corpora, the alignment between gender associations in texts and real-world demographic patterns observed here suggests that such measures can meaningfully capture gendered representations in non-Western languages and historical settings.

In summary, our findings show that female associations increased in the Work and Politics domains while the female association of the Home domain remained stable, consistent with domain-specific changes that are temporally aligned with postwar institutional transformations. 
Furthermore, the proportions of women in occupations showed positive correlations with female stereotypes, indicating that linguistic gender representations partially mirrored real-world gender representations in occupations.

\section{Limitations}
We acknowledge several limitations in our study. 
First, our training corpus may reflect a limited range of perspectives on gender norms. Metadata from the Japanese National Diet Library indicate that more than 80\% of book authors between 1922 and 1968 are inferred to be male (Appendix~\ref{secauthor}). As a result, the gender stereotypes we measure may disproportionately represent male perspectives rather than a balanced societal views. At the same time, we find that book genre distributions remain highly stable throughout this period, indicating that the observed patterns are unlikely to be driven by shifts in genre composition (Appendix~\ref{secbookcategory}). Nonetheless, the dominance of male authorship remains an important limitation when interpreting our results.
Relatedly, unlike traditional psychological approaches, our operationalization captures distributional regularities in written language and does not necessarily reflect individual-level beliefs or normative expectations.

Second, our results may be heavily influenced by the specific word lists used to represent the concepts of home, work, male, and female~\citep{ethayarajh2019understanding, antoniak2021bad}. To address this concern, we constructed the word lists based on a well-established dictionary (J-LIWC and Empath) and further refined them by removing gendered terms from the home and work domains. We also employed a bootstrap procedure to obtain stable stereotype estimates robust to variation in word list composition. It should be noted, however, that the bootstrap procedure assesses stability under resampling from the chosen word lists, and does not guarantee that alternative theoretically plausible domain definitions would produce the same results. The conclusions drawn from each domain should therefore be interpreted with this limitation in mind.

Third, since the corpus includes only up to five-grams, the maximum window size we could use for training was limited to four, which may have affected the results. Nevertheless, a window size of four is consistent with prior embedding-based studies of gender stereotypes~\citep{garg2018word, jones2020stereotypical, bhatia2021changes}, suggesting that our results remain broadly comparable to existing work.

Finally, many occupations used in the analysis had relatively few data points in the national census, making it difficult to detect statistically reliable correlations. The smaller word list for occupations, combined with the lower frequency of occupational terms, also introduces volatility in stereotype values.

\backmatter

\section*{List of abbreviations}

\begin{description}
\item[WEAT] Word Embedding Association Test
\item[ITS] Interrupted Time-Series Analysis
\item[NDL] National Diet Library
\item[LIWC] Linguistic Inquiry and Word Count
\item[J-LIWC] Japanese Linguistic Inquiry and Word Count
\item[IAT] Implicit Association Test
\end{description}

\section*{Declarations}

\begin{itemize}
\item Ethics approval and consent to participate:
Not applicable.
\item Consent for publication:
Not applicable.
\item Availability of data and materials:
The trained word embedding models are publicly available at \url{https://huggingface.co/shinsaka/japanese-word-embeddings-100/tree/main}.
The NDL Ngram Data can also be accessed at \url{https://github.com/ndl-lab/ndlngramdata}.
\item Competing interests:
The authors declare that they have no competing interests.
\item Funding:
This work was supported by JSPS KAKENHI Grant Numbers JP22K20159 and JP24K16359, and by the Research Institute of Science and Technology for Society, Japan Science and Technology Agency, Grant Number JPMJRS23L4.
\item Authors' contributions:
S.S. performed the empirical analysis.
S.S., H.K., J.A., A.M. discussed and interpreted the results and wrote the manuscript.
\item Acknowledgements:
Not applicable.
\end{itemize}

\begin{appendices}
\section{The proportion of book category for each year}\label{secbookcategory}
\begin{figure}[htbp]
\includegraphics[width=1.0\textwidth]{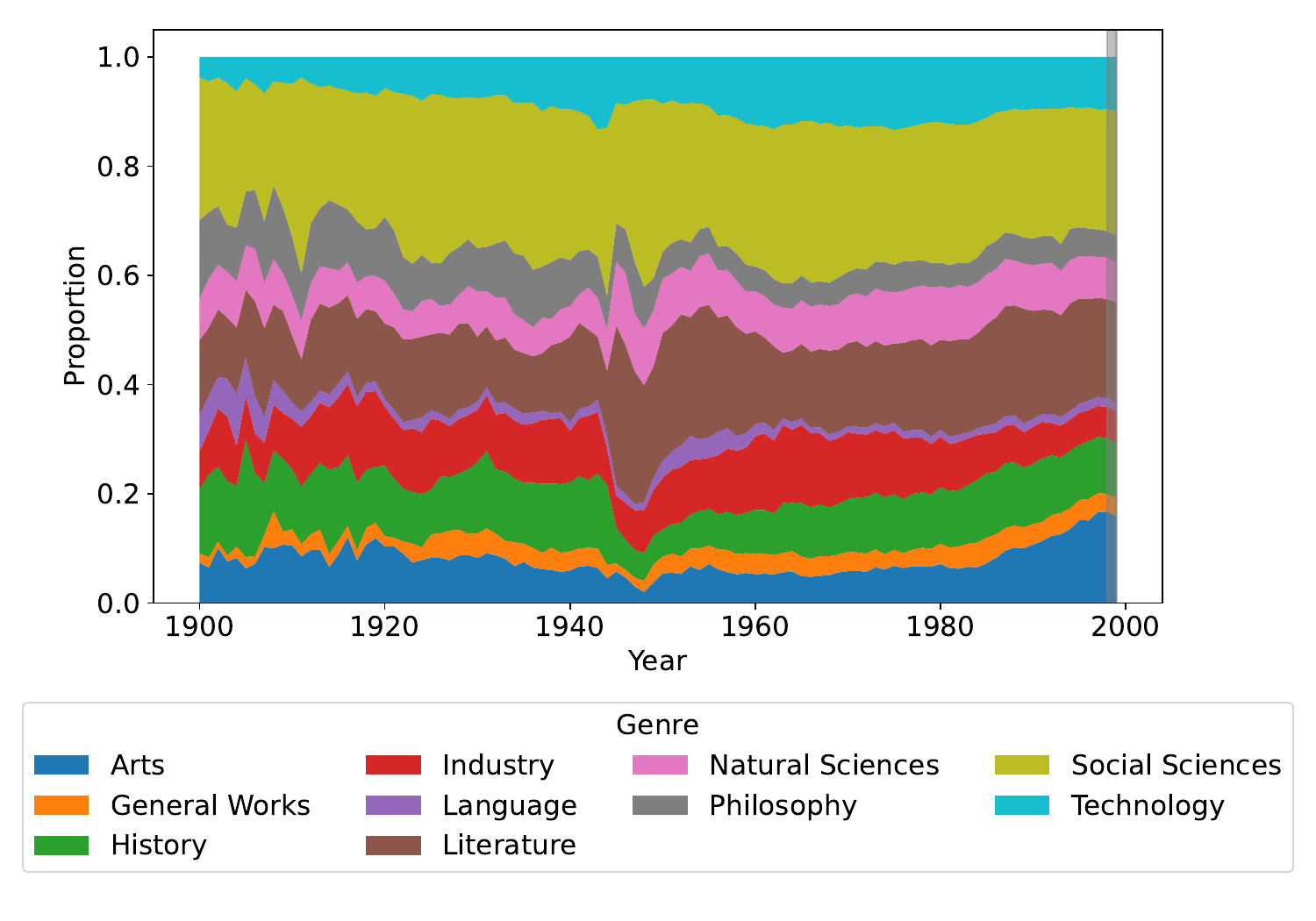}
\caption{\label{fig:book-category-composition}Temporal changes in the distribution of 10 book genres in the corpus. The shaded area represent the year 1999.}
\end{figure}

To provide a clearer overview of the corpus composition, we examined the distribution of book genres. The Japanese National Diet Library categorizes books into ten broad genres, including General Works, Philosophy, History, Social Sciences, Natural Sciences, Technology, Industry, Arts, Language, and Literature. Figure~\ref{fig:book-category-composition} illustrates how the proportions of these genres change over time. Social Sciences and Literature account for the largest shares of the corpus, followed by History and Industry. 

We further examined the temporal stability of book genre distributions. Stability was defined as the mean Spearman rank correlation between year $t$ and year $t+1$. The mean correlation was 0.966 (SD = 0.044), indicating a highly stable ranking structure over time. This high degree of stability suggests that the observed patterns are unlikely to be driven by changes in book genre composition.

\section{The inference of book authors' gender for each year}\label{secauthor}
To assess whose voices are represented in the corpus, we examined the gender composition of authors with metadata provided by the Japanese National Diet Library. We inferred authors’ gender from their names using GPT-4o mini. To validate this GPT-based name–gender inference, we evaluated the model on five publicly available Japanese name datasets~\citep{NameDataset2021,enamdict,name_origin, pham2023gendec}. Ideally, we would have constructed a validation dataset using the authors’ own names. However, most authors were insufficiently well known for reliable gender information to be obtained online, making it infeasible to build a ground-truth dataset from the corpus itself. We therefore relied on existing name datasets. From each dataset, we randomly subsampled 50 male and 50 female names, inferred gender for each name, and computed classification accuracy. Across four datasets, GPT-4o mini consistently achieved high accuracy, with an average accuracy of 0.937 (Table~\ref{tab:gpt_gender_accuracy}).

\begin{table}[ht]
\centering
\caption{GPT gender classification accuracy across four Japanese name datasets.}
\label{tab:gpt_gender_accuracy}
\begin{tabularx}{\textwidth}{l *{3}{>{\centering\arraybackslash}X}}
\hline
\textbf{Dataset} & \textbf{Accuracy} \\
\hline
enamdict          & 0.950 \\
facebook          & 0.939 \\
gendec           & 0.910 \\
name\_origin      & 0.950 \\
\hline
\textbf{Average}                        & \textbf{0.937} \\
\hline
\end{tabularx}
\end{table}

\begin{figure}[htbp]
\includegraphics[width=1.0\textwidth]{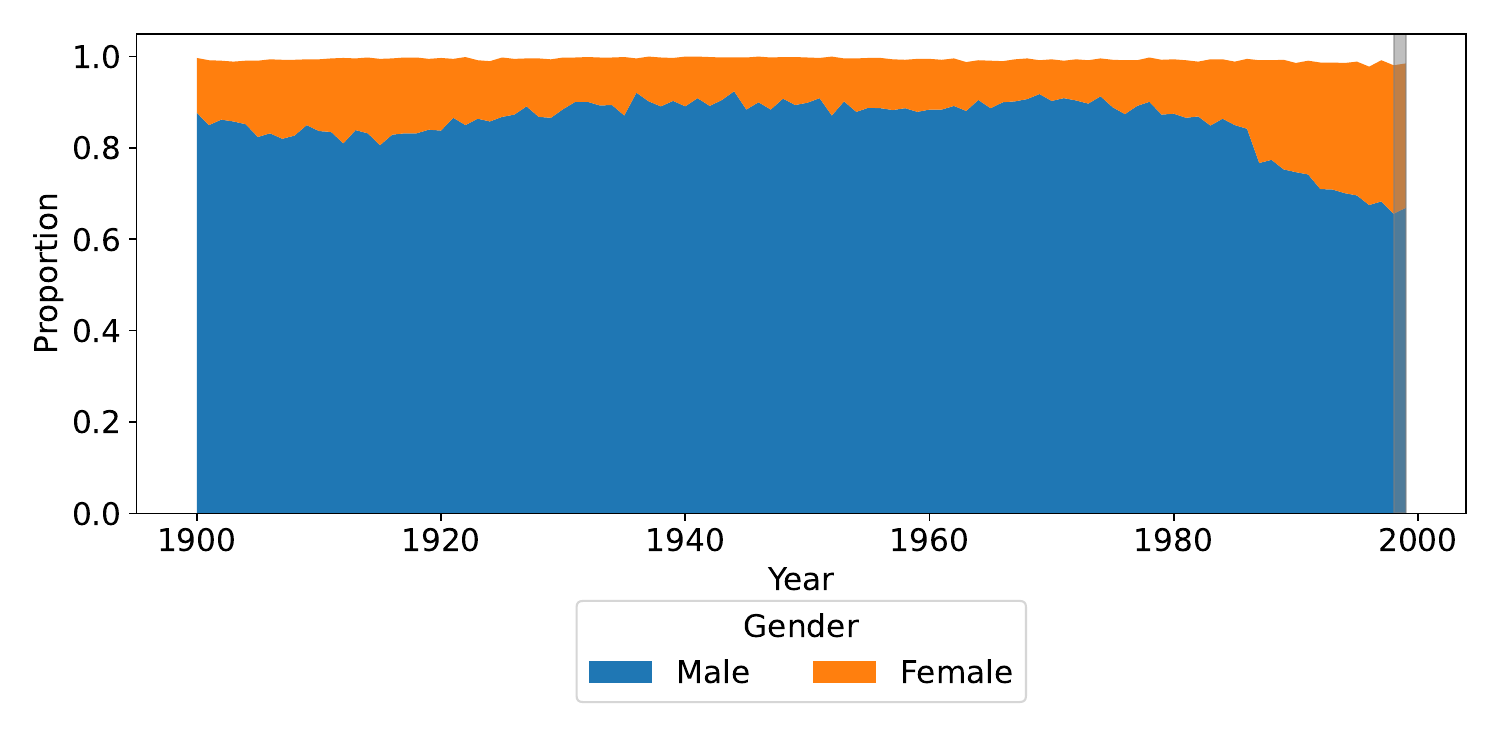}
\caption{\label{fig:author-geder-composition}Temporal trends in estimated author gender proportions (male vs. female) inferred from author names using GPT-4o mini. The shaded area represent the year 1999.}
\end{figure}

Since we found that GPT-4o mini can infer individuals’ gender from their names with high accuracy, we extracted 1,000 author names from each year and estimated the gender proportions. We find that more than 66\% (1998) to 92\% (1944) of authors between 1900 and 1999 are inferred to be male, followed by a gradual increase in female authorship in later decades (Figure~\ref{fig:author-geder-composition}).

\section{Sensitivity check: Home domain word list}\label{sec-sensitivity-home}

\begin{figure}[htbp]
\includegraphics[width=1.0\textwidth]{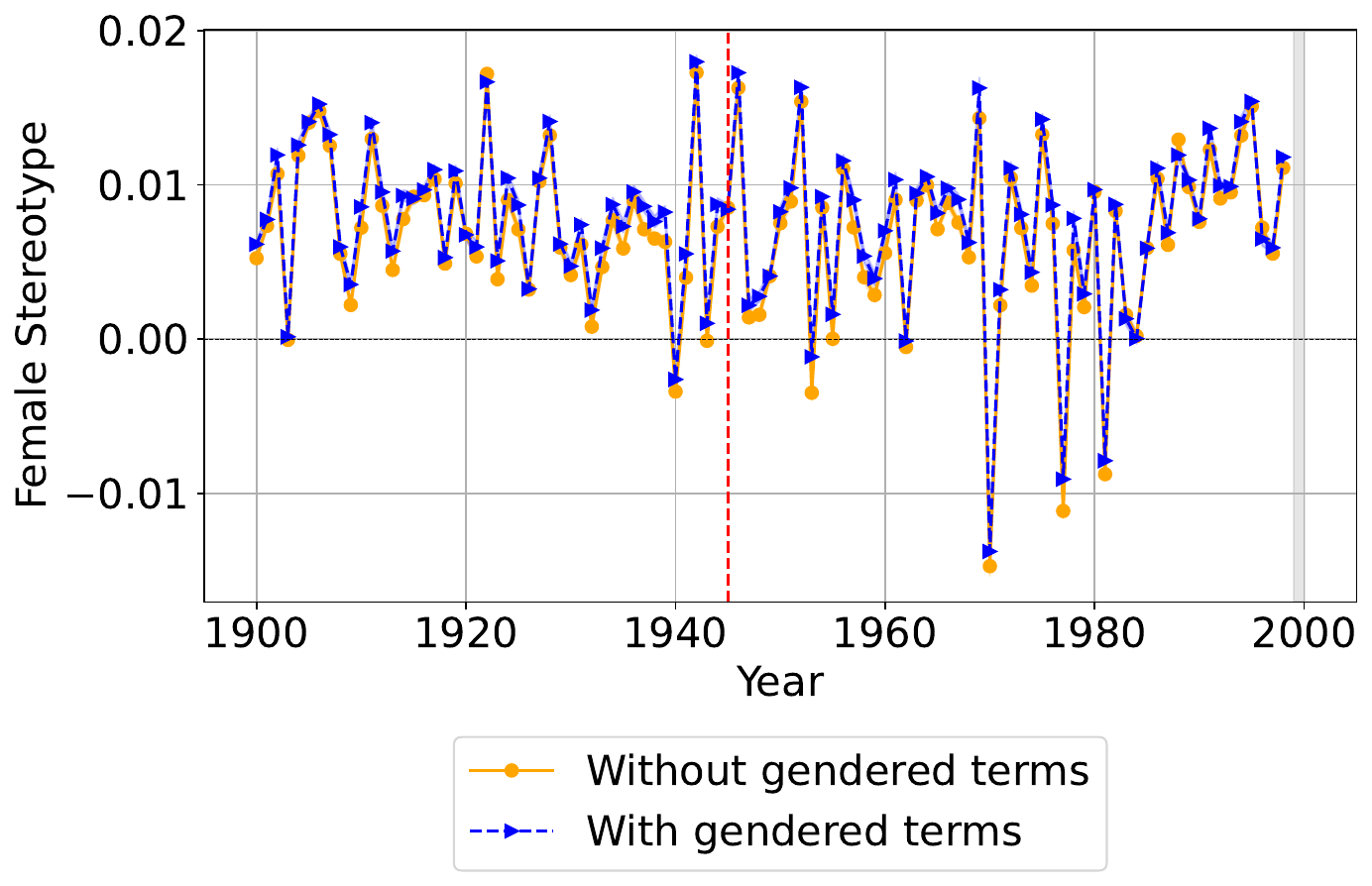}
\caption{\label{fig:sensitivity_home}\textbf{Sensitivity check for the Home domain word list.} \small Female stereotype values (WEAT scores) for the Home domain using the original word list (orange) and an extended word list including explicitly gendered domestic terms (blue). The difference between the two specifications is marginal throughout the entire period.}
\end{figure}

Figure~\ref{fig:sensitivity_home} compares female stereotype values for the Home domain using the original word list and an extended word list that additionally includes explicitly gendered domestic terms (\Ja{主婦} (housewife), \Ja{乳母} (wet nurse), \Ja{主人} (husband), \Ja{亭主} (husband), \Ja{女中} (maid)). The two specifications produce nearly identical trajectories throughout the entire 1900--1998 period, indicating that the Home domain conclusions are not materially affected by this word list choice. The average female stereotype value across 1900--1998 is 0.0068 for the original word list and 0.0077 for the extended word list. The slight increase when gendered terms are included is directionally expected: terms such as \Ja{主婦} (housewife) and \Ja{女中} (maid) are female-coded by definition, so adding them marginally raises the measured female association. Excluding them therefore yields a slightly more conservative estimate, while leaving the overall temporal trend unchanged.

\section{Alignment problem of diachironic embedding models}\label{sec-alignment}
Aligning embedding models across time is a critical step when analyzing changes in individual word vectors, because independently trained embeddings can differ by arbitrary orthogonal transformations that prevent meaningful cross-temporal comparisons~\citep{hamilton2016diachronic}. In our study, however, we compute gender association scores within each yearly embedding model, based on relative similarities between domain words and gender attribute words, and then compare these association scores across years. As discussed in the Supplementary Material of prior work~\citep{charlesworth2022historical}, we note that this type of within-period relative association is not affected by cross-temporal alignment issues.

\section{The full list of Home, Work, Politics, and Gender words}\label{sec-wordlist}
In this section, we provide the corresponding row numbers of the words in the csv dictionary file (Table~\ref{table:LIWCrownumber}), which can be accessed upon purchasing the license of LIWC2015. Since Empath is an open-source library, we provide the complete list of words used to represent the political domain (Table~\ref{table:politicswords}).

\begin{table}[htbp]
\centering
\caption{Row numbers in the CSV file for each domain}
\label{table:LIWCrownumber}
\begin{tabular}{l p{12cm}}
\hline
\textbf{Domain} & \textbf{Row numbers} \\
\hline
Home & 1650, 1712, 1780, 2282, 2443, 2553, 2670, 2710, 2996, 3007, 3026, 3045, 3168, 3242, 3292, 3300, 3534, 3736, 3878, 3879, 3880, 3882, 3883, 3884, 3886, 3887, 3888, 4028, 4029, 4054, 4101, 4213, 4220, 4225, 4837, 4924, 5281, 5471, 5506, 5780, 5781, 5782, 5783, 5785, 5788, 5791, 5792, 5843, 5844, 5845, 5846, 5964, 6026, 6157, 6158, 6688, 6689, 6691, 6714, 6973, 7138, 7140, 7141, 7241, 7440, 7478, 7606, 8084, 8435, 8498, 9032, 9106, 9228, 9340, 9673, 9816, 9948, 10427, 10557, 10639, 10645, 10935, 11057, 11180, 11205 \\
Work & 652, 653, 1973, 2361, 2399, 2444, 2461, 2487, 2543, 2544, 2606, 2655, 2667, 2684, 2713, 2740, 2872, 2957, 2977, 2979, 3012, 3032, 3093, 3131, 3138, 3165, 3212, 3213, 3370, 3371, 3374, 3418, 3451, 3480, 3564, 3572, 3622, 3630, 3643, 3680, 3693, 3735, 3736, 3737, 3738, 3797, 3802, 3804, 3805, 3808, 3809, 3810, 3820, 3829, 3831, 3832, 3834, 3835, 3837, 3944, 3947, 3968, 3990, 3991, 4020, 4030, 4033, 4039, 4052, 4088, 4102, 4103, 4104, 4108, 4117, 4120, 4121, 4146, 4182, 4200, 4202, 4244, 4246, 4255, 4256, 4313, 4371, 4387, 4413, 4432, 4434, 4452, 4461, 4478, 4481, 4483, 4491, 4577, 4591, 4592, 4594, 4601, 4604, 4605, 4606, 4620, 4621, 4630, 4648, 4654, 4655, 4689, 4691, 4715, 4716, 4718, 4720, 4734, 4791, 4792, 4797, 4833, 4836, 4856, 4873, 4875, 4887, 4962, 4963, 4964, 4986, 5013, 5016, 5106, 5135, 5136, 5137, 5138, 5139, 5140, 5141, 5142, 5146, 5172, 5173, 5174, 5175, 5177, 5178, 5236, 5268, 5277, 5280, 5291, 5292, 5293, 5331, 5343, 5345, 5346, 5482, 5507, 5527, 5529, 5554, 5632, 5633, 5686, 5687, 5688, 5689, 5690, 5691, 5692, 5695, 5696, 5697, 5698, 5699, 5700, 5701, 5702, 5703, 5705, 5706, 5708, 5709, 5742, 5762, 5763, 5764, 5767, 5769, 5770, 5782, 5790, 5802, 5851, 5876, 5879, 5936, 5937, 5939, 5944, 5946, 5976, 5996, 5998, 5999, 6024, 6032, 6033, 6034, 6046, 6080, 6081, 6086, 6089, 6095, 6097, 6105, 6106, 6136, 6163, 6164, 6171, 6187, 6264, 6266, 6267, 6310, 6352, 6353, 6381, 6392, 6410, 6584, 6662, 6665, 6668, 6669, 6706, 6742, 6745, 6761, 6763, 6766, 6783, 6784, 6801, 6843, 6844, 6901, 6904, 6907, 6909, 6911, 6912, 6913, 6941, 6979, 6992, 6997, 6998, 7000, 7001, 7002, 7003, 7004, 7005, 7008, 7022, 7029, 7063, 7074, 7115, 7127, 7129, 7130, 7142, 7164, 7165, 7166, 7168, 7189, 7190, 7191, 7192, 7194, 7197, 7199, 7217, 7219, 7248, 7249, 7275, 7291, 7304, 7305, 7359, 7360, 7361, 7362, 7430, 7470, 7471, 7472, 7476, 7477, 7478, 7479, 7480, 7481, 7532, 7558, 7573, 7588, 7622, 7626, 7657, 7658, 7660, 7661, 7666, 7667, 7668, 7669, 7691, 7692, 7699, 7701, 7708, 7712, 7723, 7726, 7733, 7735, 7736, 7749, 7835, 7891, 7957, 7988, 7990, 7993, 7994, 8046, 8048, 8064, 8226, 8289, 8301, 8347, 8398, 8437, 8533, 8565, 8576, 8579, 8582, 8601, 8632, 8698, 8706, 8710, 8719, 8720, 8758, 8760, 8764, 8765, 8766, 8769, 8770, 8776, 8807, 8817, 8827, 8914, 8920, 8932, 8934, 8968, 8970, 9011, 9023, 9036, 9037, 9042, 9062, 9064, 9065, 9067, 9136, 9150, 9163, 9167, 9181, 9182, 9189, 9190, 9191, 9192, 9245, 9258, 9282, 9295, 9297, 9316, 9336, 9338, 9339, 9359, 9360, 9361, 9362, 9376, 9377, 9378, 9389, 9390, 9401, 9408, 9427, 9484, 9491, 9492, 9533, 9534, 9535, 9536, 9537, 9538, 9569, 9591, 9615, 9617, 9840, 9877, 9903, 9957, 9961, 9962, 9963, 10057, 10058, 10059, 10120, 10127, 10152, 10154, 10156, 10162, 10163, 10164, 10166, 10167, 10171, 10172, 10177, 10183, 10184, 10185, 10189, 10193, 10198, 10199, 10200, 10214, 10236, 10279, 10280, 10283, 10285, 10291, 10297, 10298, 10299, 10304, 10305, 10318, 10320, 10325, 10326, 10343, 10345, 10347, 10348, 10350, 10353, 10370, 10372, 10390, 10424, 10429, 10433, 10436, 10437, 10439, 10440, 10449, 10499, 10571, 10572, 10696, 10697, 10703, 10747, 10748, 10749, 10750, 10787, 10820, 10822, 10832, 10895, 10914, 10919, 10932, 10938, 10939, 10940, 10989, 10995, 10997, 11011, 11022, 11048, 11060, 11073, 11074, 11089, 11124, 11125, 11159, 11187, 11200, 11203, 11205, 11267, 11269, 11272, 11315, 11318, 11348, 11351, 11352, 11418, 11420, 11486, 11528 \\
Male & 356, 3035, 4164, 4166, 5519, 5927, 6188, 6274, 6501, 8417, 8420, 8421, 8581, 8603, 8605, 8608, 8610, 8611, 8978, 9487 \\
Female & 383, 2299, 5555, 5556, 5562, 5564, 5568, 5616, 5617, 5620, 5623, 5637, 5647, 5650, 5926, 6278, 7859, 7865, 8977, 9486 \\
\hline
\end{tabular}
\end{table}

\begin{table*}[h]
\centering\small
\caption{The lists of words in the Politics domain.}
\label{table:politicswords}
\begin{tabularx}{\textwidth}{X}
\toprule
{} Politics \\
\midrule
\Ja{宣言}(declaration), \Ja{論争}(controversy), \Ja{支配}(ruling), \Ja{政治}(politics), \Ja{勅令}(decree), \Ja{議会}(parliament), \Ja{社会}(society), \Ja{外交}(diplomacy), \Ja{関与}(involvement), \Ja{国家}(nation), \Ja{運動}(campaign), \Ja{知事}(governor), \Ja{分裂}(division), \Ja{打倒}(overthrow), \Ja{陰謀}(conspiracy), \Ja{君主}(monarch), \Ja{交渉}(negotiation), \Ja{体制}(regime), \Ja{紛争}(dispute), \Ja{解放}(liberation), \Ja{政策}(policy), \Ja{州}(province), \Ja{市民}(citizen), \Ja{代表}(representative), \Ja{教義}(doctrine), \Ja{国会}(congress), \Ja{哲学}(philosophy), \Ja{合意}(consensus), \Ja{憲法}(constitution), \Ja{顧問}(advisor), \Ja{大使}(ambassador), \Ja{選ぶ}(elect), \Ja{腐敗}(corruption), \Ja{選挙}(election), \Ja{上院}(senate) \\
\bottomrule
\end{tabularx}
\end{table*}

\section{The quality of the word embedding models}\label{secA5}
\begin{figure}[htbp]
    \centering
    \begin{subfigure}[b]{0.45\textwidth}
        \centering
        \includegraphics[width=\textwidth]{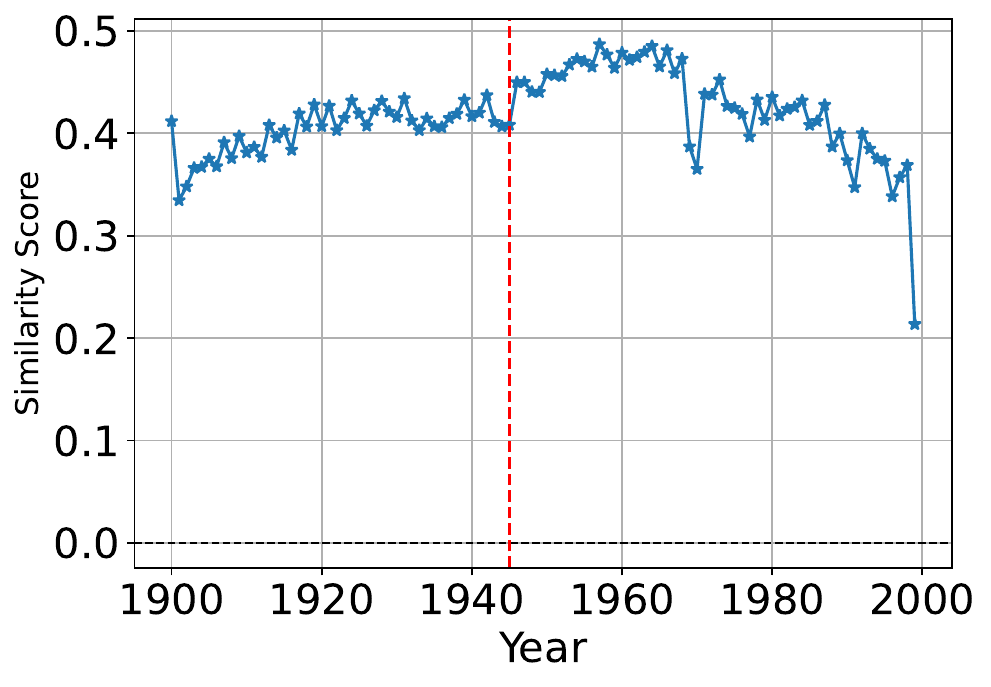}
        \caption{The similarity score across the years}
        \label{fig:quality_sim_combined}
    \end{subfigure}
    \hfill
    \begin{subfigure}[b]{0.45\textwidth}
        \centering
        \includegraphics[width=\textwidth]{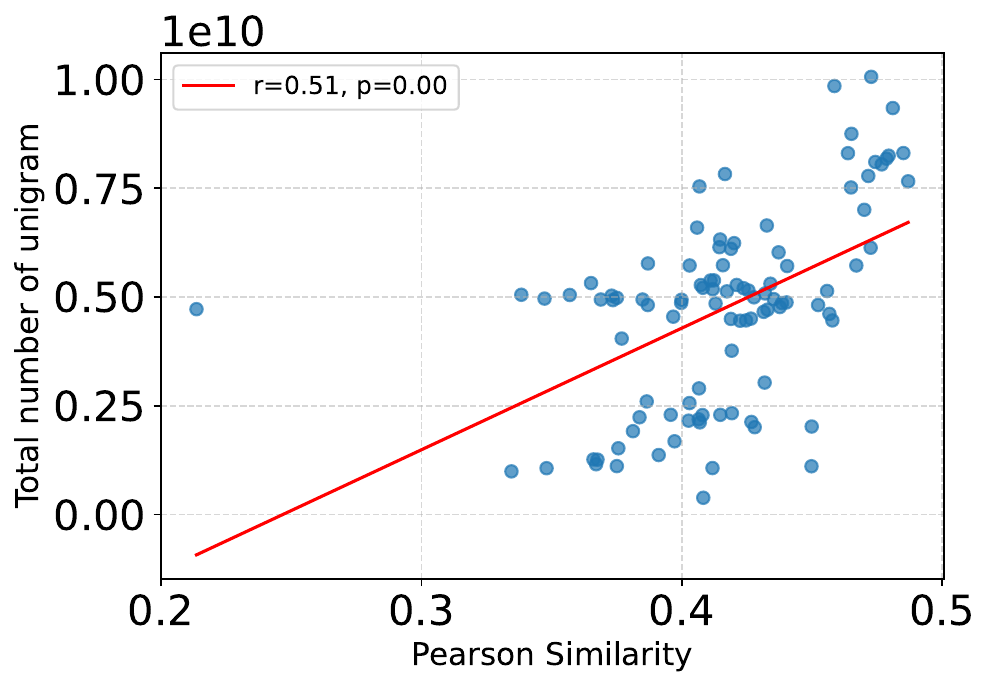}
        \caption{The correlation between similarity score and vocabulary size for each year}
        \label{fig:quality_sim_corr_combined}
    \end{subfigure}

    \vspace{0.3cm} 

    \begin{subfigure}[b]{0.45\textwidth}
        \centering
        \includegraphics[width=\textwidth]{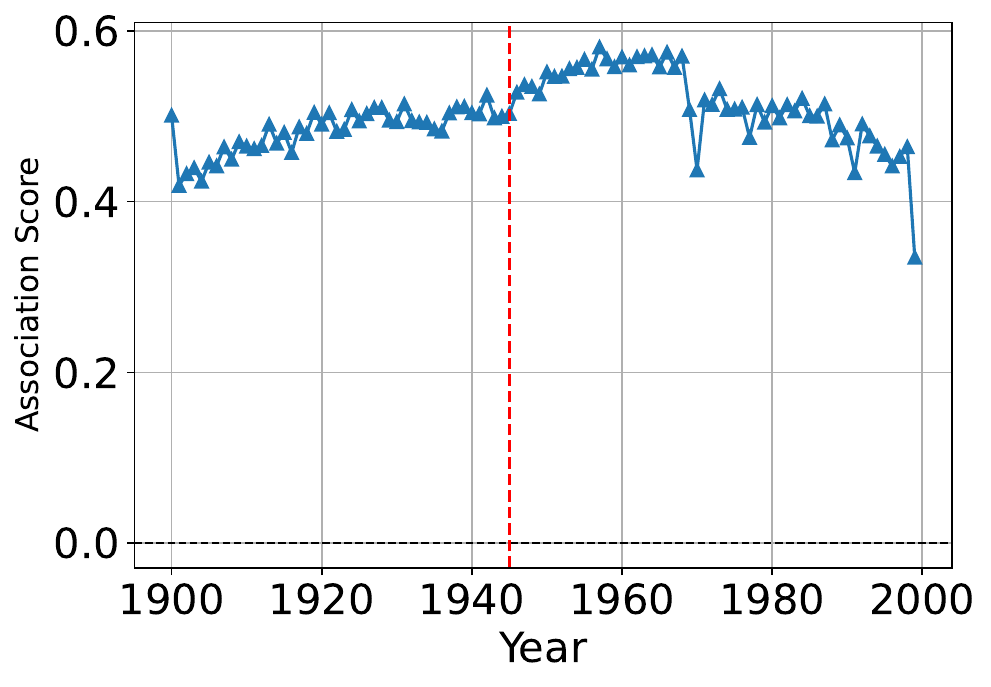}
        \caption{The association score across the year}
        \label{fig:quality_assoc_combined}
    \end{subfigure}
    \hfill
    \begin{subfigure}[b]{0.45\textwidth}
        \centering
        \includegraphics[width=\textwidth]{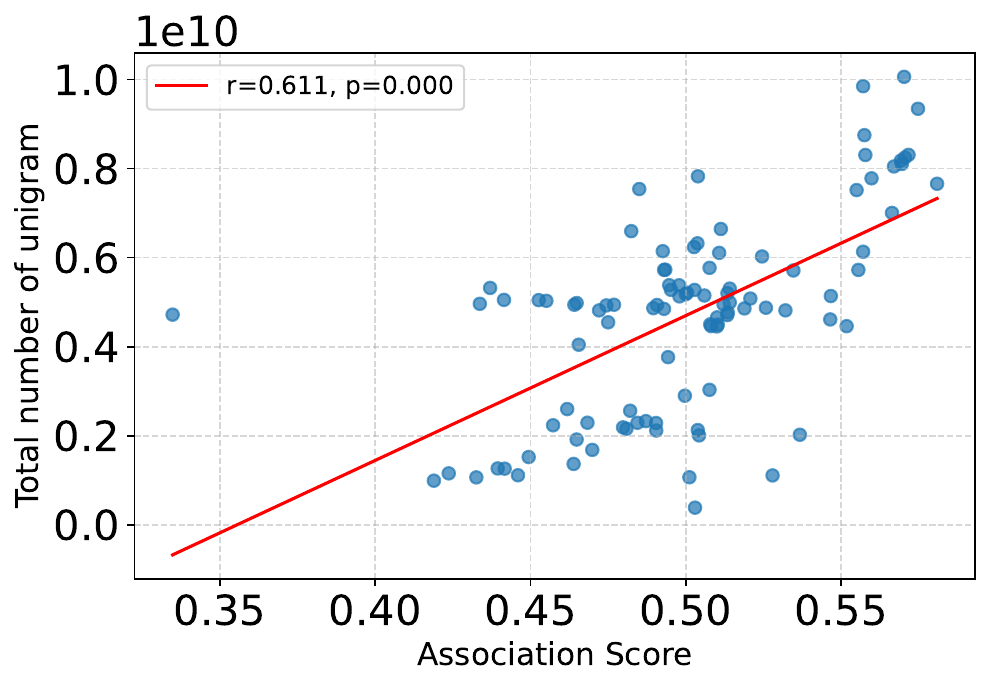}
        \caption{The correlation between association score and vocabulary size for each year}
        \label{fig:quality_assoc_corr_combined}
    \end{subfigure}

    \caption{The quality of each year's word embedding models measured by similarity and association scores, and also their correlation with vocaburary size. The red line corresponds to the year 1945. }
    \label{fig:model_quality}
\end{figure}

Figures~\ref{fig:quality_sim_combined} and~\ref{fig:quality_assoc_combined} show the yearly changes in similarity and association scores. On average, the similarity score is 0.43 and the association score is 0.47. We also find that yearly similarity and association scores are strongly correlated with the vocabulary size of the corpus (Figures~\ref{fig:quality_sim_corr_combined} and~\ref{fig:quality_assoc_corr_combined}), suggesting that vocabulary size had a substantial impact on the quality of the yearly word embedding models.

\section{The female stereotype change of individual occupations}\label{secA3}
\begin{figure}[htbp]
    \centering
    
    \begin{minipage}{0.48\textwidth}
        \centering
        \includegraphics[width=\linewidth]{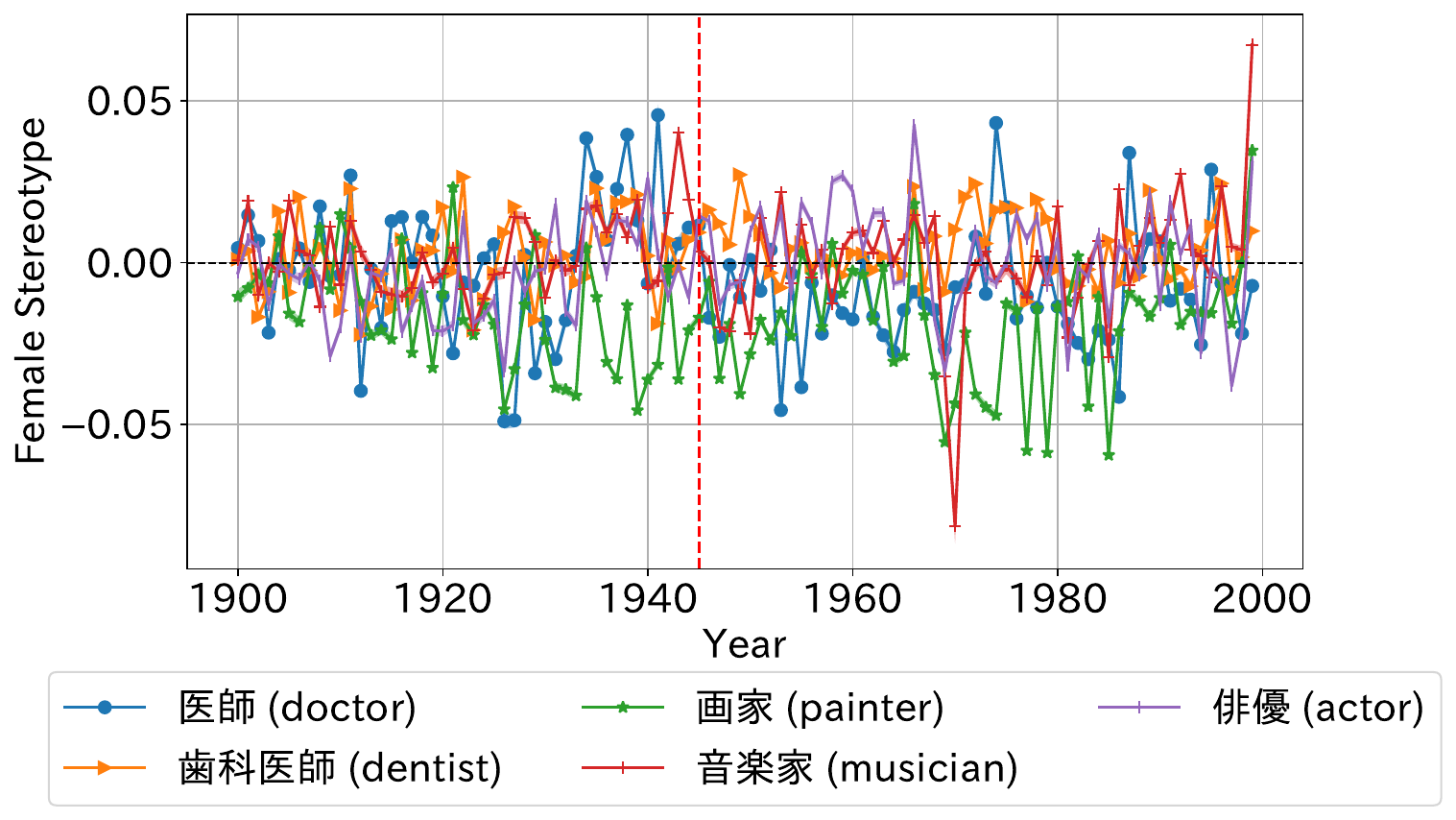}
    \end{minipage}\hfill
    \begin{minipage}{0.48\textwidth}
        \centering
        \includegraphics[width=\linewidth]{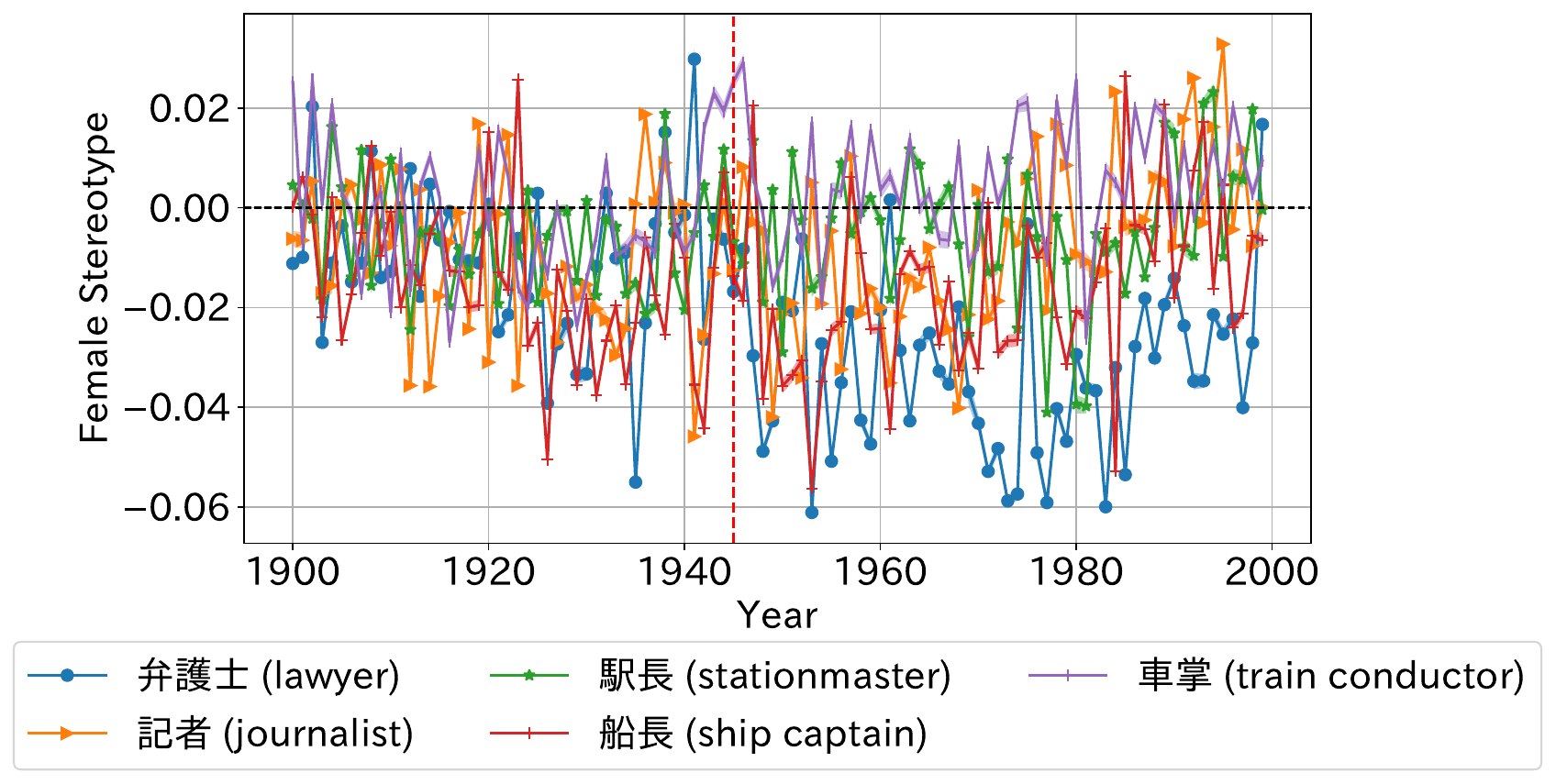}
    \end{minipage}
    
    \begin{minipage}{0.48\textwidth}
        \centering
        \includegraphics[width=\linewidth]{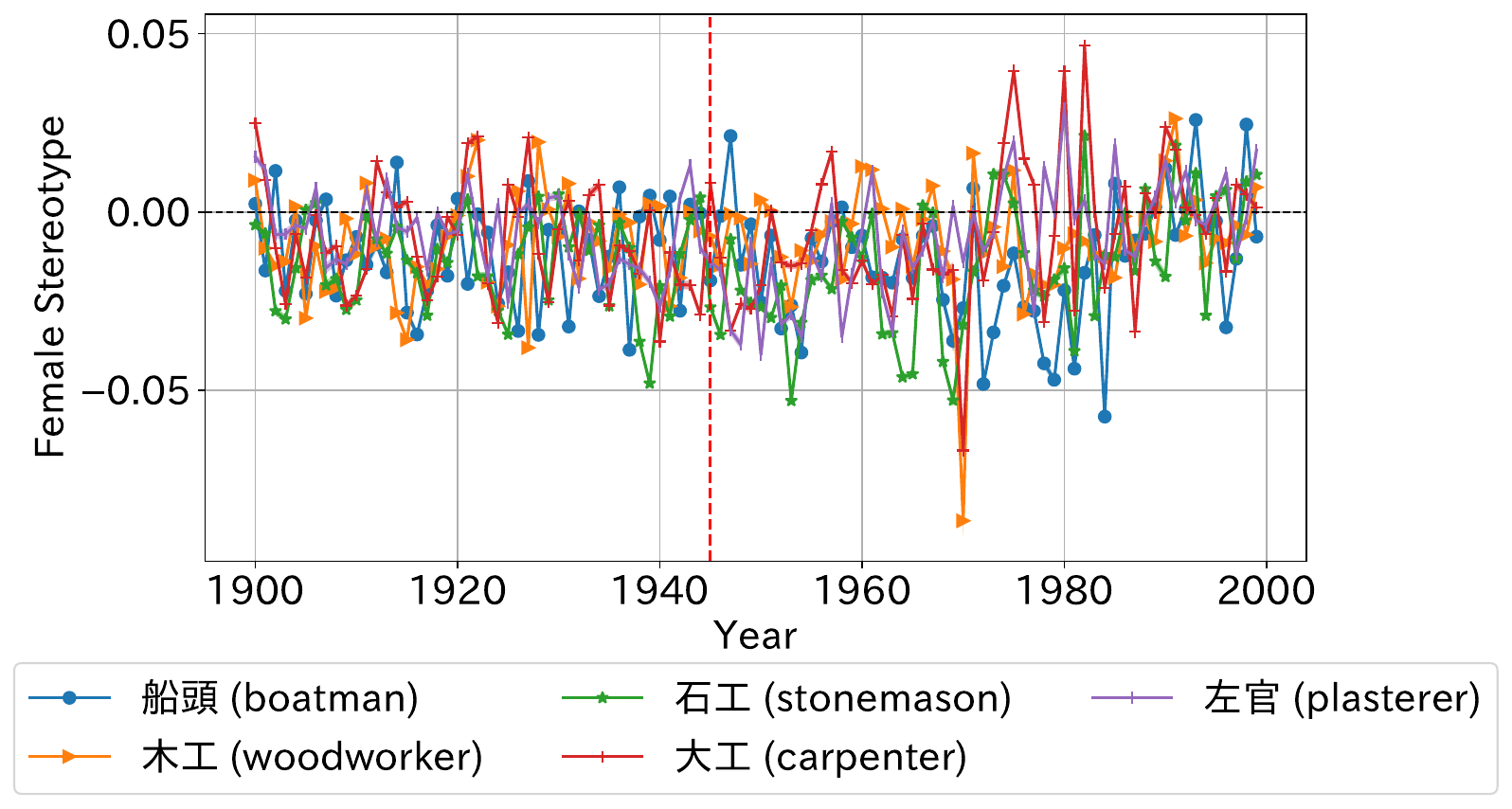}
    \end{minipage}\hfill
    \begin{minipage}{0.48\textwidth}
        \centering
        \includegraphics[width=\linewidth]{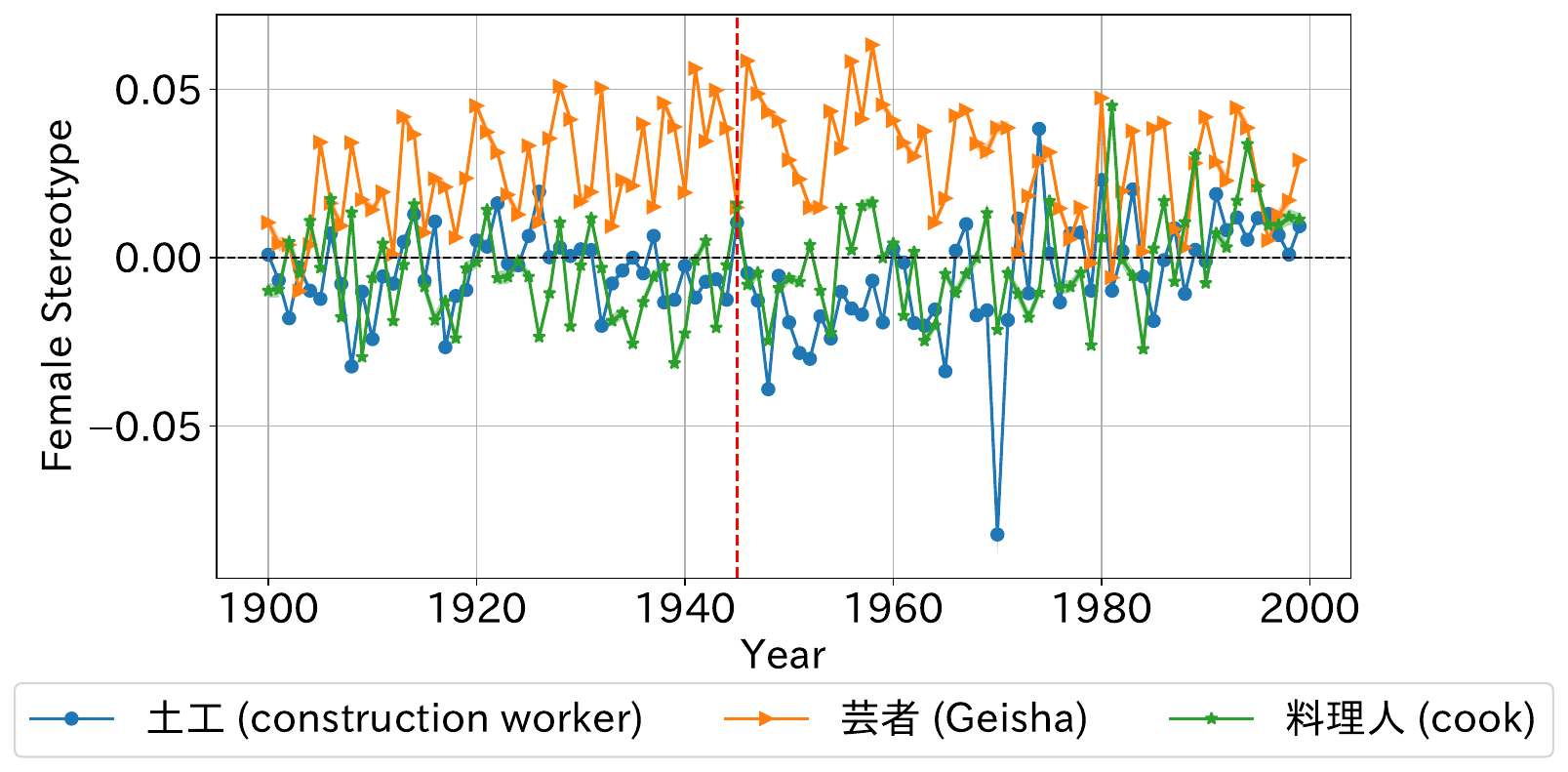}
    \end{minipage}

    \caption{\label{fig:occupations_all}The changes of female stereotypes for each occupation over the century.}
\end{figure}

Figure~\ref{fig:occupations_all} shows female stereotype change for each occupation. 

\section{Additional analyses using alternative cutoff years}\label{sec-rdd-cutoff}
This appendix presents supplemental analyses extending the interrupted time-series analysis (ITS). To assess whether the postwar trend changes reflect a sustained shift or could instead be driven by any particular postwar decade, or appear only with a temporal lag, we conducted additional analyses using a fixed 1945 cutoff while progressively varying the starting year of the post-treatment period. Specifically, the pre-period was held constant (1900–1944), and the post-period began sequentially from 1946 onward (e.g., 1946–1998, 1947–1998, and so on), with each successive specification excluding one additional year from the beginning of the post-period. For each specification, we estimated the pre-1945 trend, the post-1945 trend, and the change in trend coefficient between the two periods. Tables~\ref{tab:home_postperiod_robustness} to~\ref{tab:politics_postperiod_robustness} report the results.

The results reveal clear differences across the three domains. For the Work and Politics domains, the trend change coefficients are statistically significant for post-period specifications beginning before the mid-1980s, with p-values consistently below 0.001, while significance diminishes as the post-period is restricted to later years. Critically, significance is already present when the post-period begins in 1946, ruling out a delayed effect. Furthermore, the significance holds across a wide range of post-period specifications, indicating that these changes were not driven by any single postwar decade but instead reflect a sustained and consistent shift that persisted throughout the entire postwar period. For the Home domain, statistically significant trend changes also emerge in some post-period specifications, but the p-values are substantially larger — typically in the range of 0.01 to 0.09 — indicating considerably weaker evidence of a trend change compared to Work and Politics. Moreover, where significant, the direction of change for the Home domain is positive, meaning that home became more strongly female-associated rather than less.

The interpretation of these domain-specific patterns differs accordingly. The immediate and sustained significance for Work and Politics is temporally aligned with postwar institutional reforms, with no evidence of a delayed effect. This is consistent with the historical argument that formal reforms such as women's suffrage, labor protections, and constitutional guarantees of equality, which directly targeted participation in work and political life, corresponded to shifts in how women were represented in language in those domains. For the Home domain, the weaker and less consistent significance, combined with the direction of stronger female stereotyping, suggests that postwar reforms had limited impact on domestic gender norms in language. Two interpretations are plausible. One is ideological reinforcement: women's domestic roles may have been culturally valorized during postwar reconstruction even as women entered the workforce, amplifying the female coding of home in published texts. The other is a corpus framing effect: as women gained visibility in public life, books and magazines may have increasingly framed women in domestic contexts as a culturally distinctive role, thereby inflating the textual association between women and the home domain. We acknowledge that these remain speculative explanations, and a full account of this pattern is left to future research.

\begin{table}[htbp]
\centering
\caption{Robustness analysis for the Home domain using a fixed 1945 cutoff and varying post-treatment windows (1900--1998). $^{***}p<0.001$, $^{**}p<0.01$, $^{*}p<0.05$.}
\label{tab:home_postperiod_robustness}
\begin{tabularx}{\textwidth}{lccccc}
\toprule
Start Year & \# Post Years & Pre-trend & Post-trend & Trend Change & P-value \\
\midrule
1946 & 53 & $-8.4\times10^{-5}$ & $5.1\times10^{-5}$  & $1.35\times10^{-4}$ & 0.092 \\
1947 & 52 & $-8.4\times10^{-5}$ & $7.6\times10^{-5}$  & $1.60\times10^{-4}$ & 0.045* \\
1948 & 51 & $-8.4\times10^{-5}$ & $6.9\times10^{-5}$  & $1.53\times10^{-4}$ & 0.059 \\
1949 & 50 & $-8.4\times10^{-5}$ & $6.2\times10^{-5}$  & $1.46\times10^{-4}$ & 0.076 \\
1950 & 49 & $-8.4\times10^{-5}$ & $6.1\times10^{-5}$  & $1.45\times10^{-4}$ & 0.083 \\
1951 & 48 & $-8.4\times10^{-5}$ & $6.9\times10^{-5}$  & $1.54\times10^{-4}$ & 0.072 \\
1952 & 47 & $-8.4\times10^{-5}$ & $8.1\times10^{-5}$  & $1.65\times10^{-4}$ & 0.057 \\
1953 & 46 & $-8.4\times10^{-5}$ & $1.13\times10^{-4}$ & $1.98\times10^{-4}$ & 0.023* \\
1954 & 45 & $-8.4\times10^{-5}$ & $9.0\times10^{-5}$  & $1.75\times10^{-4}$ & 0.046* \\
1955 & 44 & $-8.4\times10^{-5}$ & $1.04\times10^{-4}$ & $1.89\times10^{-4}$ & 0.034* \\
1956 & 43 & $-8.4\times10^{-5}$ & $9.2\times10^{-5}$  & $1.77\times10^{-4}$ & 0.052 \\
1957 & 42 & $-8.4\times10^{-5}$ & $1.15\times10^{-4}$ & $1.99\times10^{-4}$ & 0.031* \\
1958 & 41 & $-8.4\times10^{-5}$ & $1.27\times10^{-4}$ & $2.11\times10^{-4}$ & 0.026* \\
1959 & 40 & $-8.4\times10^{-5}$ & $1.28\times10^{-4}$ & $2.13\times10^{-4}$ & 0.029* \\
1960 & 39 & $-8.4\times10^{-5}$ & $1.25\times10^{-4}$ & $2.09\times10^{-4}$ & 0.037* \\
1961 & 38 & $-8.4\times10^{-5}$ & $1.31\times10^{-4}$ & $2.15\times10^{-4}$ & 0.037* \\
1962 & 37 & $-8.4\times10^{-5}$ & $1.54\times10^{-4}$ & $2.38\times10^{-4}$ & 0.025* \\
1963 & 36 & $-8.4\times10^{-5}$ & $1.34\times10^{-4}$ & $2.18\times10^{-4}$ & 0.045* \\
1964 & 35 & $-8.4\times10^{-5}$ & $1.58\times10^{-4}$ & $2.42\times10^{-4}$ & 0.031* \\
1965 & 34 & $-8.4\times10^{-5}$ & $1.92\times10^{-4}$ & $2.76\times10^{-4}$ & 0.017* \\
1966 & 33 & $-8.4\times10^{-5}$ & $2.16\times10^{-4}$ & $3.01\times10^{-4}$ & 0.012* \\
1967 & 32 & $-8.4\times10^{-5}$ & $2.49\times10^{-4}$ & $3.34\times10^{-4}$ & 0.007** \\
1968 & 31 & $-8.4\times10^{-5}$ & $2.86\times10^{-4}$ & $3.71\times10^{-4}$ & 0.004** \\
1969 & 30 & $-8.4\times10^{-5}$ & $3.13\times10^{-4}$ & $3.98\times10^{-4}$ & 0.003** \\
1970 & 29 & $-8.4\times10^{-5}$ & $4.03\times10^{-4}$ & $4.87\times10^{-4}$ & $4.0\times10^{-4}$*** \\
1971 & 28 & $-8.4\times10^{-5}$ & $2.83\times10^{-4}$ & $3.68\times10^{-4}$ & 0.005** \\
1972 & 27 & $-8.4\times10^{-5}$ & $2.71\times10^{-4}$ & $3.55\times10^{-4}$ & 0.010* \\
1973 & 26 & $-8.4\times10^{-5}$ & $3.38\times10^{-4}$ & $4.23\times10^{-4}$ & 0.003** \\
1974 & 25 & $-8.4\times10^{-5}$ & $3.79\times10^{-4}$ & $4.63\times10^{-4}$ & 0.002** \\
1975 & 24 & $-8.4\times10^{-5}$ & $3.93\times10^{-4}$ & $4.77\times10^{-4}$ & 0.003** \\
1976 & 23 & $-8.4\times10^{-5}$ & $5.27\times10^{-4}$ & $6.12\times10^{-4}$ & $3.0\times10^{-4}$*** \\
1977 & 22 & $-8.4\times10^{-5}$ & $6.17\times10^{-4}$ & $7.01\times10^{-4}$ & $7.0\times10^{-5}$*** \\
1978 & 21 & $-8.4\times10^{-5}$ & $4.59\times10^{-4}$ & $5.44\times10^{-4}$ & 0.002** \\
1979 & 20 & $-8.4\times10^{-5}$ & $5.03\times10^{-4}$ & $5.88\times10^{-4}$ & 0.002** \\
1980 & 19 & $-8.4\times10^{-5}$ & $4.91\times10^{-4}$ & $5.76\times10^{-4}$ & 0.004** \\
1981 & 18 & $-8.4\times10^{-5}$ & $6.26\times10^{-4}$ & $7.11\times10^{-4}$ & 0.001** \\
1982 & 17 & $-8.4\times10^{-5}$ & $3.85\times10^{-4}$ & $4.70\times10^{-4}$ & 0.030* \\
1983 & 16 & $-8.4\times10^{-5}$ & $4.57\times10^{-4}$ & $5.42\times10^{-4}$ & 0.023* \\
1984 & 15 & $-8.4\times10^{-5}$ & $3.47\times10^{-4}$ & $4.31\times10^{-4}$ & 0.094 \\
1985 & 14 & $-8.4\times10^{-5}$ & $1.19\times10^{-4}$ & $2.04\times10^{-4}$ & 0.459 \\
1986 & 13 & $-8.4\times10^{-5}$ & $2.1\times10^{-5}$  & $1.05\times10^{-4}$ & 0.732 \\
1987 & 12 & $-8.4\times10^{-5}$ & $6.6\times10^{-5}$  & $1.51\times10^{-4}$ & 0.666 \\
1988 & 11 & $-8.4\times10^{-5}$ & $-1.06\times10^{-4}$& $-2.1\times10^{-5}$ & 0.957 \\
1989 & 10 & $-8.4\times10^{-5}$ & $-2.5\times10^{-5}$ & $5.9\times10^{-5}$  & 0.898 \\
1990 & 9  & $-8.4\times10^{-5}$ & $-7.6\times10^{-5}$ & $8.0\times10^{-6}$  & 0.988 \\
\bottomrule
\end{tabularx}
\end{table}

\begin{table}[htbp]
\centering
\caption{Robustness analysis for the Work domain using a fixed 1945 cutoff and varying post-treatment windows (1900--1998). $^{***}p<0.001$, $^{**}p<0.01$, $^{*}p<0.05$.}
\label{tab:work_postperiod_robustness}
\begin{tabularx}{\textwidth}{lccccc}
\toprule
Start Year & \# Post Years & Pre-trend & Post-trend & Trend Change & P-value \\
\midrule
1946 & 53 & $-1.7\times10^{-4}$ & $1.57\times10^{-4}$ & $3.26\times10^{-4}$ & $3.5\times10^{-5}$*** \\
1947 & 52 & $-1.7\times10^{-4}$ & $1.69\times10^{-4}$ & $3.38\times10^{-4}$ & $2.3\times10^{-5}$*** \\
1948 & 51 & $-1.7\times10^{-4}$ & $1.49\times10^{-4}$ & $3.19\times10^{-4}$ & $6.2\times10^{-5}$*** \\
1949 & 50 & $-1.7\times10^{-4}$ & $1.53\times10^{-4}$ & $3.23\times10^{-4}$ & $6.7\times10^{-5}$*** \\
1950 & 49 & $-1.7\times10^{-4}$ & $1.66\times10^{-4}$ & $3.36\times10^{-4}$ & $4.4\times10^{-5}$*** \\
1951 & 48 & $-1.7\times10^{-4}$ & $1.71\times10^{-4}$ & $3.40\times10^{-4}$ & $4.9\times10^{-5}$*** \\
1952 & 47 & $-1.7\times10^{-4}$ & $1.90\times10^{-4}$ & $3.60\times10^{-4}$ & $2.3\times10^{-5}$*** \\
1953 & 46 & $-1.7\times10^{-4}$ & $2.13\times10^{-4}$ & $3.82\times10^{-4}$ & $9.0\times10^{-6}$*** \\
1954 & 45 & $-1.7\times10^{-4}$ & $2.12\times10^{-4}$ & $3.82\times10^{-4}$ & $1.4\times10^{-5}$*** \\
1955 & 44 & $-1.7\times10^{-4}$ & $2.37\times10^{-4}$ & $4.07\times10^{-4}$ & $5.0\times10^{-6}$*** \\
1956 & 43 & $-1.7\times10^{-4}$ & $2.12\times10^{-4}$ & $3.82\times10^{-4}$ & $1.9\times10^{-5}$*** \\
1957 & 42 & $-1.7\times10^{-4}$ & $2.15\times10^{-4}$ & $3.85\times10^{-4}$ & $2.5\times10^{-5}$*** \\
1958 & 41 & $-1.7\times10^{-4}$ & $2.46\times10^{-4}$ & $4.16\times10^{-4}$ & $7.5\times10^{-6}$*** \\
1959 & 40 & $-1.7\times10^{-4}$ & $2.48\times10^{-4}$ & $4.18\times10^{-4}$ & $1.1\times10^{-5}$*** \\
1960 & 39 & $-1.7\times10^{-4}$ & $2.45\times10^{-4}$ & $4.14\times10^{-4}$ & $2.2\times10^{-5}$*** \\
1961 & 38 & $-1.7\times10^{-4}$ & $2.69\times10^{-4}$ & $4.38\times10^{-4}$ & $1.2\times10^{-5}$*** \\
1962 & 37 & $-1.7\times10^{-4}$ & $2.83\times10^{-4}$ & $4.53\times10^{-4}$ & $1.2\times10^{-5}$*** \\
1963 & 36 & $-1.7\times10^{-4}$ & $2.83\times10^{-4}$ & $4.53\times10^{-4}$ & $2.1\times10^{-5}$*** \\
1964 & 35 & $-1.7\times10^{-4}$ & $3.14\times10^{-4}$ & $4.83\times10^{-4}$ & $1.1\times10^{-5}$*** \\
1965 & 34 & $-1.7\times10^{-4}$ & $3.37\times10^{-4}$ & $5.07\times10^{-4}$ & $8.0\times10^{-6}$*** \\
1966 & 33 & $-1.7\times10^{-4}$ & $3.61\times10^{-4}$ & $5.31\times10^{-4}$ & $6.4\times10^{-6}$*** \\
1967 & 32 & $-1.7\times10^{-4}$ & $4.34\times10^{-4}$ & $6.03\times10^{-4}$ & $4.4\times10^{-7}$*** \\
1968 & 31 & $-1.7\times10^{-4}$ & $4.62\times10^{-4}$ & $6.32\times10^{-4}$ & $3.6\times10^{-7}$*** \\
1969 & 30 & $-1.7\times10^{-4}$ & $4.68\times10^{-4}$ & $6.37\times10^{-4}$ & $8.1\times10^{-7}$*** \\
1970 & 29 & $-1.7\times10^{-4}$ & $4.23\times10^{-4}$ & $5.93\times10^{-4}$ & $6.9\times10^{-6}$*** \\
1971 & 28 & $-1.7\times10^{-4}$ & $3.42\times10^{-4}$ & $5.11\times10^{-4}$ & $9.5\times10^{-5}$*** \\
1972 & 27 & $-1.7\times10^{-4}$ & $3.25\times10^{-4}$ & $4.95\times10^{-4}$ & $3.0\times10^{-4}$*** \\
1973 & 26 & $-1.7\times10^{-4}$ & $3.84\times10^{-4}$ & $5.53\times10^{-4}$ & $1.1\times10^{-4}$*** \\
1974 & 25 & $-1.7\times10^{-4}$ & $4.05\times10^{-4}$ & $5.75\times10^{-4}$ & $1.5\times10^{-4}$*** \\
1975 & 24 & $-1.7\times10^{-4}$ & $4.46\times10^{-4}$ & $6.16\times10^{-4}$ & $1.2\times10^{-4}$*** \\
1976 & 23 & $-1.7\times10^{-4}$ & $5.27\times10^{-4}$ & $6.97\times10^{-4}$ & $3.8\times10^{-5}$*** \\
1977 & 22 & $-1.7\times10^{-4}$ & $6.35\times10^{-4}$ & $8.04\times10^{-4}$ & $6.5\times10^{-6}$*** \\
1978 & 21 & $-1.7\times10^{-4}$ & $5.46\times10^{-4}$ & $7.15\times10^{-4}$ & $1.0\times10^{-4}$*** \\
1979 & 20 & $-1.7\times10^{-4}$ & $5.82\times10^{-4}$ & $7.52\times10^{-4}$ & $1.5\times10^{-4}$*** \\
1980 & 19 & $-1.7\times10^{-4}$ & $6.07\times10^{-4}$ & $7.76\times10^{-4}$ & $2.7\times10^{-4}$*** \\
1981 & 18 & $-1.7\times10^{-4}$ & $6.92\times10^{-4}$ & $8.62\times10^{-4}$ & $1.9\times10^{-4}$*** \\
1982 & 17 & $-1.7\times10^{-4}$ & $5.25\times10^{-4}$ & $6.94\times10^{-4}$ & 0.003706** \\
1983 & 16 & $-1.7\times10^{-4}$ & $5.14\times10^{-4}$ & $6.83\times10^{-4}$ & 0.009001** \\
1984 & 15 & $-1.7\times10^{-4}$ & $4.31\times10^{-4}$ & $6.00\times10^{-4}$ & 0.035* \\
1985 & 14 & $-1.7\times10^{-4}$ & $2.68\times10^{-4}$ & $4.38\times10^{-4}$ & 0.159 \\
1986 & 13 & $-1.7\times10^{-4}$ & $-4.1\times10^{-5}$ & $1.29\times10^{-4}$ & 0.699 \\
1987 & 12 & $-1.7\times10^{-4}$ & $-2.11\times10^{-4}$ & $-4.1\times10^{-5}$ & 0.912 \\
1988 & 11 & $-1.7\times10^{-4}$ & $-4.80\times10^{-4}$ & $-3.10\times10^{-4}$ & 0.465 \\
1989 & 10 & $-1.7\times10^{-4}$ & $-5.56\times10^{-4}$ & $-3.86\times10^{-4}$ & 0.434 \\
1990 & 9  & $-1.7\times10^{-4}$ & $-4.20\times10^{-4}$ & $-2.50\times10^{-4}$ & 0.667 \\
\bottomrule
\end{tabularx}
\end{table}

\begin{table}[htbp]
\centering
\caption{Robustness analysis for the Politics domain using a fixed 1945 cutoff and varying post-treatment windows (1900--1998). $^{***}p<0.001$, $^{**}p<0.01$, $^{*}p<0.05$.}
\label{tab:politics_postperiod_robustness}
\begin{tabularx}{\textwidth}{lccccc}
\toprule
Start Year & \# Post Years & Pre-trend & Post-trend & Trend Change & P-value \\
\midrule
1946 & 53 & $-2.74\times10^{-4}$ & $2.26\times10^{-4}$ & $5.00\times10^{-4}$ & $4.0\times10^{-6}$*** \\
1947 & 52 & $-2.74\times10^{-4}$ & $2.31\times10^{-4}$ & $5.06\times10^{-4}$ & $4.2\times10^{-6}$*** \\
1948 & 51 & $-2.74\times10^{-4}$ & $2.01\times10^{-4}$ & $4.76\times10^{-4}$ & $1.3\times10^{-5}$*** \\
1949 & 50 & $-2.74\times10^{-4}$ & $2.10\times10^{-4}$ & $4.85\times10^{-4}$ & $1.3\times10^{-5}$*** \\
1950 & 49 & $-2.74\times10^{-4}$ & $2.21\times10^{-4}$ & $4.96\times10^{-4}$ & $1.1\times10^{-5}$*** \\
1951 & 48 & $-2.74\times10^{-4}$ & $2.33\times10^{-4}$ & $5.07\times10^{-4}$ & $1.0\times10^{-5}$*** \\
1952 & 47 & $-2.74\times10^{-4}$ & $2.64\times10^{-4}$ & $5.38\times10^{-4}$ & $4.0\times10^{-6}$*** \\
1953 & 46 & $-2.74\times10^{-4}$ & $2.90\times10^{-4}$ & $5.65\times10^{-4}$ & $1.7\times10^{-6}$*** \\
1954 & 45 & $-2.74\times10^{-4}$ & $2.91\times10^{-4}$ & $5.65\times10^{-4}$ & $2.7\times10^{-6}$*** \\
1955 & 44 & $-2.74\times10^{-4}$ & $3.32\times10^{-4}$ & $6.06\times10^{-4}$ & $6.2\times10^{-7}$*** \\
1956 & 43 & $-2.74\times10^{-4}$ & $3.15\times10^{-4}$ & $5.89\times10^{-4}$ & $1.8\times10^{-6}$*** \\
1957 & 42 & $-2.74\times10^{-4}$ & $3.37\times10^{-4}$ & $6.12\times10^{-4}$ & $1.3\times10^{-6}$*** \\
1958 & 41 & $-2.74\times10^{-4}$ & $3.45\times10^{-4}$ & $6.20\times10^{-4}$ & $1.7\times10^{-6}$*** \\
1959 & 40 & $-2.74\times10^{-4}$ & $3.61\times10^{-4}$ & $6.35\times10^{-4}$ & $1.7\times10^{-6}$*** \\
1960 & 39 & $-2.74\times10^{-4}$ & $3.61\times10^{-4}$ & $6.35\times10^{-4}$ & $3.2\times10^{-6}$*** \\
1961 & 38 & $-2.74\times10^{-4}$ & $4.07\times10^{-4}$ & $6.82\times10^{-4}$ & $9.8\times10^{-7}$*** \\
1962 & 37 & $-2.74\times10^{-4}$ & $4.14\times10^{-4}$ & $6.88\times10^{-4}$ & $1.6\times10^{-6}$*** \\
1963 & 36 & $-2.74\times10^{-4}$ & $4.13\times10^{-4}$ & $6.87\times10^{-4}$ & $3.3\times10^{-6}$*** \\
1964 & 35 & $-2.74\times10^{-4}$ & $4.44\times10^{-4}$ & $7.19\times10^{-4}$ & $2.4\times10^{-6}$*** \\
1965 & 34 & $-2.74\times10^{-4}$ & $4.67\times10^{-4}$ & $7.41\times10^{-4}$ & $2.7\times10^{-6}$*** \\
1966 & 33 & $-2.74\times10^{-4}$ & $5.09\times10^{-4}$ & $7.83\times10^{-4}$ & $1.6\times10^{-6}$*** \\
1967 & 32 & $-2.74\times10^{-4}$ & $5.94\times10^{-4}$ & $8.68\times10^{-4}$ & $2.1\times10^{-7}$*** \\
1968 & 31 & $-2.74\times10^{-4}$ & $6.57\times10^{-4}$ & $9.31\times10^{-4}$ & $7.6\times10^{-8}$*** \\
1969 & 30 & $-2.74\times10^{-4}$ & $6.65\times10^{-4}$ & $9.39\times10^{-4}$ & $1.8\times10^{-7}$*** \\
1970 & 29 & $-2.74\times10^{-4}$ & $5.21\times10^{-4}$ & $7.95\times10^{-4}$ & $3.8\times10^{-6}$*** \\
1971 & 28 & $-2.74\times10^{-4}$ & $3.93\times10^{-4}$ & $6.67\times10^{-4}$ & $6.8\times10^{-5}$*** \\
1972 & 27 & $-2.74\times10^{-4}$ & $3.51\times10^{-4}$ & $6.25\times10^{-4}$ & $3.2\times10^{-4}$*** \\
1973 & 26 & $-2.74\times10^{-4}$ & $4.22\times10^{-4}$ & $6.96\times10^{-4}$ & $1.3\times10^{-4}$*** \\
1974 & 25 & $-2.74\times10^{-4}$ & $4.89\times10^{-4}$ & $7.64\times10^{-4}$ & $6.9\times10^{-5}$*** \\
1975 & 24 & $-2.74\times10^{-4}$ & $5.23\times10^{-4}$ & $7.97\times10^{-4}$ & $8.7\times10^{-5}$*** \\
1976 & 23 & $-2.74\times10^{-4}$ & $6.29\times10^{-4}$ & $9.04\times10^{-4}$ & $2.5\times10^{-5}$*** \\
1977 & 22 & $-2.74\times10^{-4}$ & $7.77\times10^{-4}$ & $1.05\times10^{-3}$ & $3.1\times10^{-6}$*** \\
1978 & 21 & $-2.74\times10^{-4}$ & $6.13\times10^{-4}$ & $8.87\times10^{-4}$ & $9.1\times10^{-5}$*** \\
1979 & 20 & $-2.74\times10^{-4}$ & $6.15\times10^{-4}$ & $8.89\times10^{-4}$ & $2.5\times10^{-4}$*** \\
1980 & 19 & $-2.74\times10^{-4}$ & $6.39\times10^{-4}$ & $9.13\times10^{-4}$ & $4.8\times10^{-4}$*** \\
1981 & 18 & $-2.74\times10^{-4}$ & $7.23\times10^{-4}$ & $9.98\times10^{-4}$ & $4.3\times10^{-4}$*** \\
1982 & 17 & $-2.74\times10^{-4}$ & $5.99\times10^{-4}$ & $8.73\times10^{-4}$ & 0.003784** \\
1983 & 16 & $-2.74\times10^{-4}$ & $6.56\times10^{-4}$ & $9.30\times10^{-4}$ & 0.004921** \\
1984 & 15 & $-2.74\times10^{-4}$ & $5.11\times10^{-4}$ & $7.85\times10^{-4}$ & 0.028* \\
1985 & 14 & $-2.74\times10^{-4}$ & $3.90\times10^{-4}$ & $6.64\times10^{-4}$ & 0.092 \\
1986 & 13 & $-2.74\times10^{-4}$ & $-5.0\times10^{-5}$ & $2.24\times10^{-4}$ & 0.590 \\
1987 & 12 & $-2.74\times10^{-4}$ & $-2.39\times10^{-4}$ & $3.5\times10^{-5}$ & 0.941 \\
1988 & 11 & $-2.74\times10^{-4}$ & $-5.13\times10^{-4}$ & $-2.39\times10^{-4}$ & 0.654 \\
1989 & 10 & $-2.74\times10^{-4}$ & $-5.41\times10^{-4}$ & $-2.67\times10^{-4}$ & 0.667 \\
1990 & 9  & $-2.74\times10^{-4}$ & $-3.03\times10^{-4}$ & $-2.9\times10^{-5}$ & 0.968 \\
\bottomrule
\end{tabularx}
\end{table}

\section{The quality of book-only models}\label{sec-book-analysis}
\begin{figure}[htbp]
\centering

\begin{subfigure}[t]{0.48\textwidth}
    \centering
    \includegraphics[width=\textwidth]{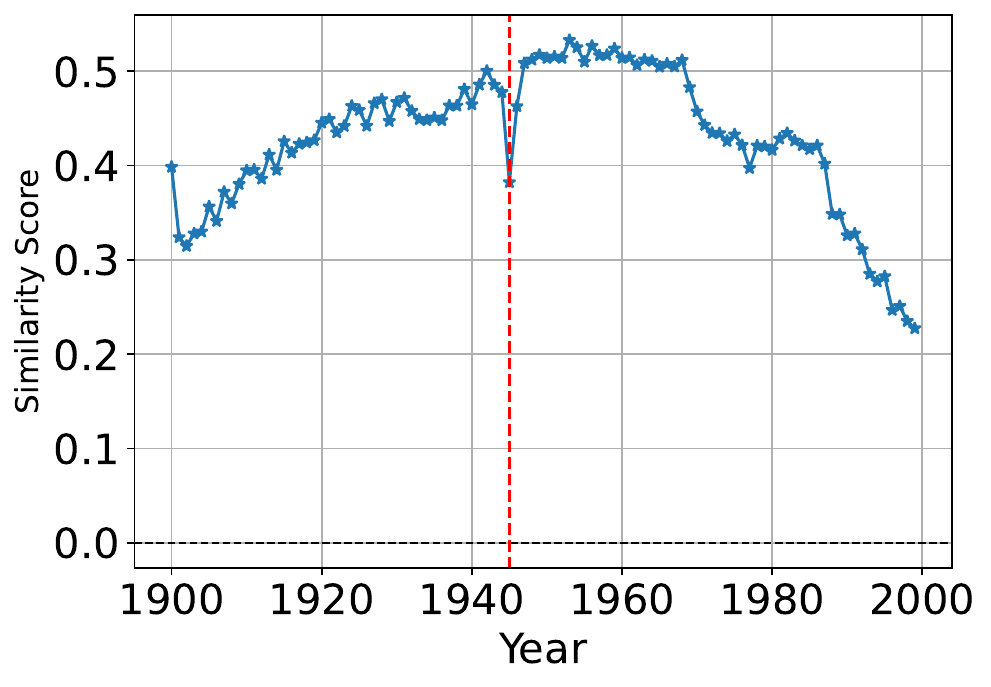}
    \caption{Similarity scores}
    \label{fig:Similarity}
\end{subfigure}
\hfill
\begin{subfigure}[t]{0.48\textwidth}
    \centering
    \includegraphics[width=\textwidth]{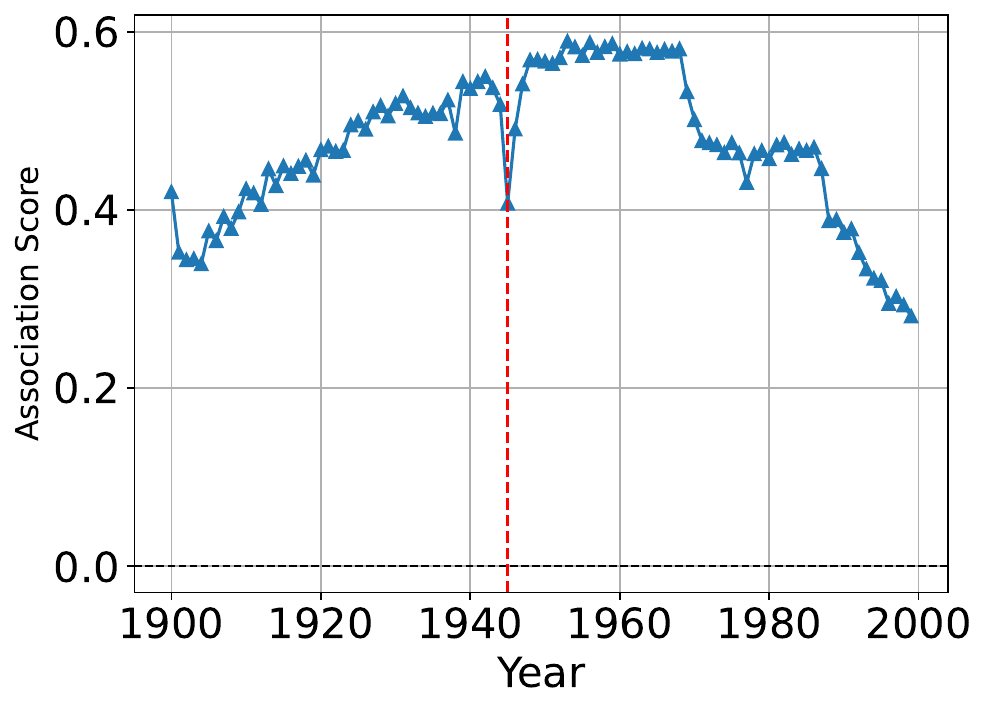}
    \caption{Association scores}
    \label{fig:Association}
\end{subfigure}

\caption{The quality of each year’s word embedding models trained on book data, measured by similarity and association scores. The red line corresponds to the year 1945.}
\label{fig:quality-scores-book}
\end{figure}

We evaluated the quality of book-only models trained for each year. Figure~\ref{fig:quality-scores-book} shows the yearly changes in similarity and association scores. On average, the similarity score is 0.43 and the association score is 0.47. However, the quality scores exhibit substantial variation across years over the 100-year period, making reliable inference difficult for many years. Therefore, we exclude the book-only models from our main analysis. Nevertheless, for transparency and reproducibility, we have uploaded the book-only models to Hugging Face (\href{https://huggingface.co/shinsaka/japanese-word-embeddings-100/tree/main}{Hugging Face repository}).

\section{The quality of magazine-only models}\label{sec-magazine-analysis}

\begin{figure}[htbp]
\centering

\begin{subfigure}[t]{0.48\textwidth}
    \centering
    \includegraphics[width=\textwidth]{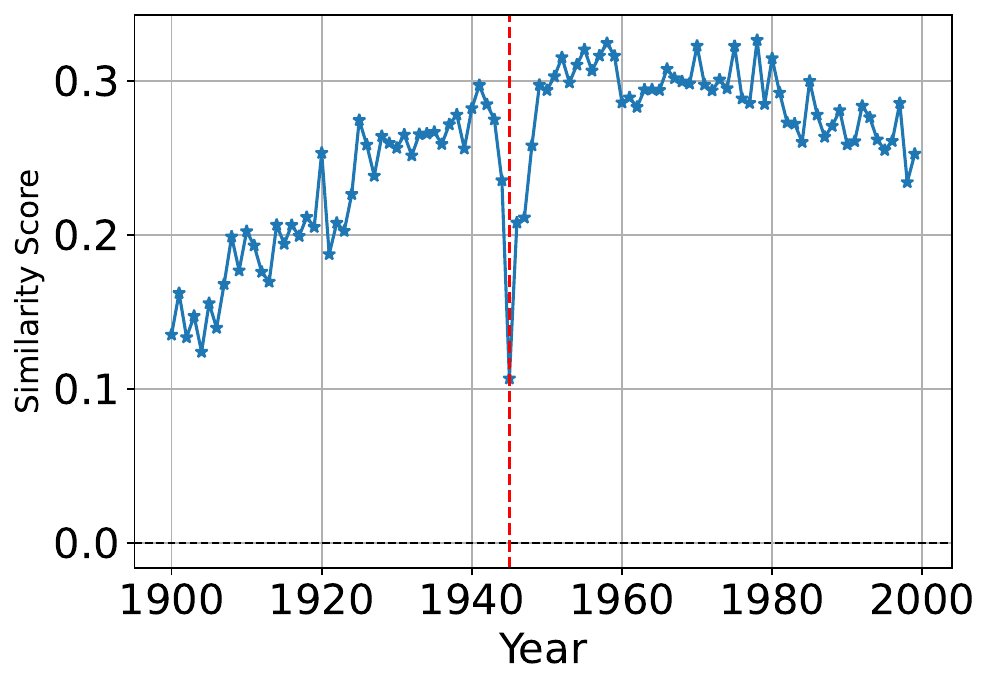}
    \caption{Similarity scores}
    \label{fig:Similarity-mag}
\end{subfigure}
\hfill
\begin{subfigure}[t]{0.48\textwidth}
    \centering
    \includegraphics[width=\textwidth]{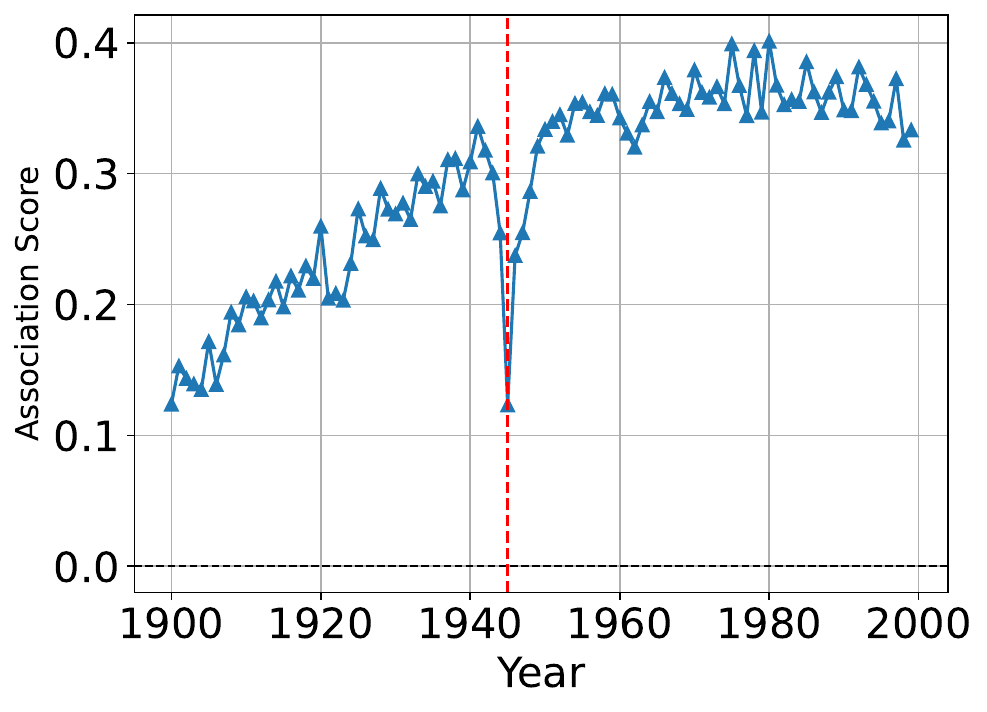}
    \caption{Association scores}
    \label{fig:Association-mag}
\end{subfigure}

\caption{The quality of each year’s word embedding models trained on magazine data, measured by similarity and association scores. The red line corresponds to the year 1945.}
\label{fig:quality-scores-mag}
\end{figure}

We evaluated the quality of magazine-only models trained for each year. Figure~\ref{fig:quality-scores-mag} shows the yearly changes in similarity and association scores. On average, the similarity score is 0.25 and the association score is 0.29. These values are substantially lower than those obtained from the combined models (0.43 and 0.47), indicating that the magazine-only corpus alone is insufficient to produce reliable embeddings. Therefore, we exclude the magazine-only models from our main analysis. Nevertheless, for transparency and reproducibility, we have uploaded the magazine-only models to Hugging Face (\href{https://huggingface.co/shinsaka/japanese-word-embeddings-100/tree/main}{Hugging Face repository}).

\end{appendices}

\nocite{*}  
\bibliography{sn-bibliography}

@book{mackie1997creating,
  title={Creating Socialist Women in Japan: Gender, Labour and Activism, 1900--1937},
  author={Mackie, Vera},
  year={1997},
  publisher={Cambridge University Press},
  address={Cambridge}
}

@book{shillony1991politics,
  title={Politics and culture in wartime Japan},
  author={Shillony, Ben-Ami},
  year={1991},
  publisher={Oxford University Press},
  address={Oxford}
}

@article{caliskan2017semantics,
  title={Semantics derived automatically from language corpora contain human-like biases},
  author={Caliskan, Aylin and Bryson, Joanna J and Narayanan, Arvind},
  journal={Science},
  volume={356},
  number={6334},
  pages={183--186},
  year={2017},
  publisher={American Association for the Advancement of Science}
}

@article{garg2018word,
  title={Word embeddings quantify 100 years of gender and ethnic stereotypes},
  author={Garg, Nikhil and Schiebinger, Londa and Jurafsky, Dan and Zou, James},
  journal={Proceedings of the National Academy of Sciences},
  volume={115},
  number={16},
  pages={E3635--E3644},
  year={2018},
  publisher={National Acad Sciences}
}

@article{kozlowski2019geometry,
  title={The geometry of culture: Analyzing the meanings of class through word embeddings},
  author={Kozlowski, Austin C and Taddy, Matt and Evans, James A},
  journal={American Sociological Review},
  volume={84},
  number={5},
  pages={905--949},
  year={2019},
  publisher={SAGE Publications Sage CA: Los Angeles, CA}
}

@article{lewis2020gender,
  title={Gender stereotypes are reflected in the distributional structure of 25 languages},
  author={Lewis, Molly and Lupyan, Gary},
  journal={Nature human behaviour},
  volume={4},
  number={10},
  pages={1021--1028},
  year={2020},
  publisher={Nature Publishing Group UK London}
}

@article{charlesworth2021gender,
  title={Gender stereotypes in natural language: Word embeddings show robust consistency across child and adult language corpora of more than 65 million words},
  author={Charlesworth, Tessa ES and Yang, Victor and Mann, Thomas C and Kurdi, Benedek and Banaji, Mahzarin R},
  journal={Psychological Science},
  volume={32},
  number={2},
  pages={218--240},
  year={2021},
  publisher={Sage Publications Sage CA: Los Angeles, CA}
}

@article{kurdi2019relationship,
  title={The relationship between implicit intergroup attitudes and beliefs},
  author={Kurdi, Benedek and Mann, Thomas C and Charlesworth, Tessa ES and Banaji, Mahzarin R},
  journal={Proceedings of the National Academy of Sciences},
  volume={116},
  number={13},
  pages={5862--5871},
  year={2019},
  publisher={National Acad Sciences}
}

@article{jones2020stereotypical,
  title={Stereotypical gender associations in language have decreased over time},
  author={Jones, Jason J and Amin, Mohammad Ruhul and Kim, Jessica and Skiena, Steven},
  journal={Sociological Science},
  volume={7},
  pages={1--35},
  year={2020}
}

@article{diekman2000stereotypes,
  title={Stereotypes as dynamic constructs: Women and men of the past, present, and future},
  author={Diekman, Amanda B and Eagly, Alice H},
  journal={Personality and social psychology bulletin},
  volume={26},
  number={10},
  pages={1171--1188},
  year={2000},
  publisher={Sage Publications Sage CA: Thousand Oaks, CA}
}

@article{donnelly2017masculine,
  title={Masculine and feminine traits on the Bem Sex-Role Inventory, 1993--2012: A cross-temporal meta-analysis},
  author={Donnelly, Kristin and Twenge, Jean M},
  journal={Sex roles},
  volume={76},
  pages={556--565},
  year={2017},
  publisher={Springer}
}

@article{eagly2020gender,
  title={Gender stereotypes have changed: A cross-temporal meta-analysis of US public opinion polls from 1946 to 2018.},
  author={Eagly, Alice H and Nater, Christa and Miller, David I and Kaufmann, Mich{\`e}le and Sczesny, Sabine},
  journal={American psychologist},
  volume={75},
  number={3},
  pages={301},
  year={2020},
  publisher={American Psychological Association}
}

@article{bolukbasi2016man,
  title={Man is to computer programmer as woman is to homemaker? debiasing word embeddings},
  author={Bolukbasi, Tolga and Chang, Kai-Wei and Zou, James Y and Saligrama, Venkatesh and Kalai, Adam T},
  journal={Advances in neural information processing systems},
  volume={29},
  year={2016}
}

@inproceedings{zhao2017ngram2vec,
  title={Ngram2vec: Learning improved word representations from ngram co-occurrence statistics},
  author={Zhao, Zhe and Liu, Tao and Li, Shen and Li, Bofang and Du, Xiaoyong},
  booktitle={Proceedings of the 2017 conference on empirical methods in natural language processing},
  pages={244--253},
  year={2017}
}

@inproceedings{antoniak2021bad,
  title={Bad seeds: Evaluating lexical methods for bias measurement},
  author={Antoniak, Maria and Mimno, David},
  booktitle={Proceedings of the 59th Annual Meeting of the Association for Computational Linguistics and the 11th International Joint Conference on Natural Language Processing (Volume 1: Long Papers)},
  pages={1889--1904},
  year={2021}
}

@inproceedings{mikolov2013linguistic,
  title={Linguistic regularities in continuous space word representations},
  author={Mikolov, Tom{\'a}{\v{s}} and Yih, Wen-tau and Zweig, Geoffrey},
  booktitle={Proceedings of the 2013 conference of the north american chapter of the association for computational linguistics: Human language technologies},
  pages={746--751},
  year={2013}
}

@inproceedings{zhao2018gender,
  title={Gender Bias in Coreference Resolution: Evaluation and Debiasing Methods},
  author={Zhao, Jieyu and Wang, Tianlu and Yatskar, Mark and Ordonez, Vicente and Chang, Kai-Wei},
  booktitle={Proceedings of the 2018 Conference of the North American Chapter of the Association for Computational Linguistics: Human Language Technologies, Volume 2 (Short Papers)},
  pages={15--20},
  year={2018}
}

@article{greenwald1998measuring,
  title={Measuring individual differences in implicit cognition: the implicit association test.},
  author={Greenwald, Anthony G and McGhee, Debbie E and Schwartz, Jordan LK},
  journal={Journal of personality and social psychology},
  volume={74},
  number={6},
  pages={1464},
  year={1998},
  publisher={American Psychological Association}
}

@inproceedings{zhao2018learning,
  title={Learning Gender-Neutral Word Embeddings},
  author={Zhao, Jieyu and Zhou, Yichao and Li, Zeyu and Wang, Wei and Chang, Kai-Wei},
  booktitle={Proceedings of the 2018 Conference on Empirical Methods in Natural Language Processing},
  pages={4847--4853},
  year={2018}
}

@article{firth1957synopsis,
  title={A synopsis of linguistic theory, 1930-1955},
  author={Firth, John},
  journal={Studies in linguistic analysis},
  pages={10--32},
  year={1957}
}

@article{gilbert1951stereotype,
  title={Stereotype persistence and change among college students.},
  author={Gilbert, Gustave M},
  journal={The Journal of Abnormal and Social Psychology},
  volume={46},
  number={2},
  pages={245},
  year={1951},
  publisher={American Psychological Association}
}

@article{karlins1969fading,
  title={On the fading of social stereotypes: studies in three generations of college students.},
  author={Karlins, Marvin and Coffman, Thomas L and Walters, Gary},
  journal={Journal of personality and social psychology},
  volume={13},
  number={1},
  pages={1},
  year={1969},
  publisher={American Psychological Association}
}

@article{bergsieker2012stereotyping,
  title={Stereotyping by omission: eliminate the negative, accentuate the positive.},
  author={Bergsieker, Hilary B and Leslie, Lisa M and Constantine, Vanessa S and Fiske, Susan T},
  journal={Journal of personality and social psychology},
  volume={102},
  number={6},
  pages={1214},
  year={2012},
  publisher={American Psychological Association}
}

@article{nosek2009national,
  title={National differences in gender--science stereotypes predict national sex differences in science and math achievement},
  author={Nosek, Brian A and Smyth, Frederick L and Sriram, Natarajan and Lindner, Nicole M and Devos, Thierry and Ayala, Alfonso and Bar-Anan, Yoav and Bergh, Robin and Cai, Huajian and Gonsalkorale, Karen and others},
  journal={Proceedings of the National Academy of Sciences},
  volume={106},
  number={26},
  pages={10593--10597},
  year={2009},
  publisher={National Acad Sciences}
}

@article{kudo2005mecab,
  title={Mecab: Yet another part-of-speech and morphological analyzer},
  author={Kudo, Taku},
  journal={http://mecab. sourceforge. net/},
  year={2005}
}

@article{ogiso2013morphological,
  title   = {Morphological analysis of historical Japanese text},
  author  = {Ogiso, Toshinobu and Komachi, Mamoru and Matsumoto, Yuji},
  journal = {Journal of Natural Language Processing},
  volume  = {20},
  number  = {5},
  pages   = {727--748},
  year    = {2013},
}

@article{prates2020assessing,
  title={Assessing gender bias in machine translation: a case study with google translate},
  author={Prates, Marcelo OR and Avelar, Pedro H and Lamb, Lu{\'\i}s C},
  journal={Neural Computing and Applications},
  volume={32},
  pages={6363--6381},
  year={2020},
  publisher={Springer}
}

@article{black2020ai,
  title={AI-enabled recruiting: What is it and how should a manager use it?},
  author={Black, J Stewart and van Esch, Patrick},
  journal={Business Horizons},
  volume={63},
  number={2},
  pages={215--226},
  year={2020},
  publisher={Elsevier}
}

@article{katz1933racial,
  title={Racial stereotypes of one hundred college students.},
  author={Katz, Daniel and Braly, Kenneth},
  journal={The Journal of Abnormal and Social Psychology},
  volume={28},
  number={3},
  pages={280},
  year={1933},
  publisher={American Psychological Association}
}

@article{devine1995racial,
  title={Are racial stereotypes really fading? The Princeton trilogy revisited},
  author={Devine, Patricia G and Elliot, Andrew J},
  journal={Personality and social psychology bulletin},
  volume={21},
  number={11},
  pages={1139--1150},
  year={1995},
  publisher={Sage Publications Sage CA: Thousand Oaks, CA}
}

@article{madon2001ethnic,
  title={Ethnic and national stereotypes: The Princeton trilogy revisited and revised},
  author={Madon, Stephanie and Guyll, Max and Aboufadel, Kathy and Montiel, Eulices and Smith, Alison and Palumbo, Polly and Jussim, Lee},
  journal={Personality and social psychology bulletin},
  volume={27},
  number={8},
  pages={996--1010},
  year={2001},
  publisher={Sage Publications Sage CA: Thousand Oaks, CA}
}

@article{williams1977sex,
  title={Sex stereotypes and trait favorability on the Adjective Check List},
  author={Williams, John E and Best, Deborah L},
  journal={Educational and psychological measurement},
  volume={37},
  number={1},
  pages={101--110},
  year={1977},
  publisher={Sage Publications Sage CA: Thousand Oaks, CA}
}

@article{williams1990measuring,
  title={Measuring sex stereotypes: A multination study, Rev},
  author={Williams, John E and Best, Deborah L},
  year={1990},
  journal={Sage Publications, Inc}
}

@article{cherlin1981trends,
  title={Trends in United States men's and women's sex-role attitudes: 1972 to 1978},
  author={Cherlin, Andrew and Walters, Pamela Barnhouse},
  journal={American Sociological Review},
  pages={453--460},
  year={1981},
  publisher={JSTOR}
}

@article{baron2006development,
  title={The development of implicit attitudes: Evidence of race evaluations from ages 6 and 10 and adulthood},
  author={Baron, Andrew Scott and Banaji, Mahzarin R},
  journal={Psychological science},
  volume={17},
  number={1},
  pages={53--58},
  year={2006},
  publisher={SAGE Publications Sage CA: Los Angeles, CA}
}

@article{ziegert2005employment,
  title={Employment discrimination: the role of implicit attitudes, motivation, and a climate for racial bias.},
  author={Ziegert, Jonathan C and Hanges, Paul J},
  journal={Journal of applied psychology},
  volume={90},
  number={3},
  pages={553},
  year={2005},
  publisher={American Psychological Association}
}

@article{banse2001implicit,
  title={Implicit attitudes towards homosexuality: Reliability, validity, and controllability of the IAT},
  author={Banse, Rainer and Seise, Jan and Zerbes, Nikola},
  journal={Zeitschrift f{\"u}r experimentelle Psychologie},
  volume={48},
  number={2},
  pages={145--160},
  year={2001}
}

@article{jellison2004implicit,
  title={Implicit and explicit measures of sexual orientation attitudes: In group preferences and related behaviors and beliefs among gay and straight men},
  author={Jellison, William A and McConnell, Allen R and Gabriel, Shira},
  journal={Personality and Social Psychology Bulletin},
  volume={30},
  number={5},
  pages={629--642},
  year={2004},
  publisher={Sage Publications}
}

@article{kiefer2007implicit,
  title={Implicit stereotypes and women’s math performance: How implicit gender-math stereotypes influence women’s susceptibility to stereotype threat},
  author={Kiefer, Amy K and Sekaquaptewa, Denise},
  journal={Journal of experimental social psychology},
  volume={43},
  number={5},
  pages={825--832},
  year={2007},
  publisher={Elsevier}
}

@article{nosek2002harvesting,
  title={Harvesting implicit group attitudes and beliefs from a demonstration web site.},
  author={Nosek, Brian A and Banaji, Mahzarin R and Greenwald, Anthony G},
  journal={Group Dynamics: Theory, research, and practice},
  volume={6},
  number={1},
  pages={101},
  year={2002},
  publisher={Educational Publishing Foundation}
}

@article{greenwald2002unified,
  title={A unified theory of implicit attitudes, stereotypes, self-esteem, and self-concept.},
  author={Greenwald, Anthony G and Banaji, Mahzarin R and Rudman, Laurie A and Farnham, Shelly D and Nosek, Brian A and Mellott, Deborah S},
  journal={Psychological review},
  volume={109},
  number={1},
  pages={3},
  year={2002},
  publisher={American Psychological Association}
}

@article{chen2023ethics,
  title={Ethics and discrimination in artificial intelligence-enabled recruitment practices},
  author={Chen, Zhisheng},
  journal={Humanities and Social Sciences Communications},
  volume={10},
  number={1},
  pages={1--12},
  year={2023},
  publisher={Palgrave}
}

@inproceedings{font2019equalizing,
  title={Equalizing Gender Bias in Neural Machine Translation with Word Embeddings Techniques},
  author={Font, Joel Escud{\'e} and Costa-juss{\`a}, Marta R},
  booktitle={Proceedings of the First Workshop on Gender Bias in Natural Language Processing},
  pages={147--154},
  year={2019}
}

@article{wang2020double,
  title={Double-hard debias: Tailoring word embeddings for gender bias mitigation},
  author={Wang, Tianlu and Lin, Xi Victoria and Rajani, Nazneen Fatema and McCann, Bryan and Ordonez, Vicente and Xiong, Caiming},
  journal={arXiv preprint arXiv:2005.00965},
  year={2020}
}

@article{bhatia2017semantic,
  title={The semantic representation of prejudice and stereotypes},
  author={Bhatia, Sudeep},
  journal={Cognition},
  volume={164},
  pages={46--60},
  year={2017},
  publisher={Elsevier}
}

@inproceedings{hamilton2016diachronic,
  title={Diachronic Word Embeddings Reveal Statistical Laws of Semantic Change},
  author={Hamilton, William L and Leskovec, Jure and Jurafsky, Dan},
  booktitle={Proceedings of the 54th Annual Meeting of the Association for Computational Linguistics (Volume 1: Long Papers)},
  pages={1489--1501},
  year={2016}
}

@article{bhatia2021changes,
  title={Changes in gender stereotypes over time: A computational analysis},
  author={Bhatia, Nazl{\i} and Bhatia, Sudeep},
  journal={Psychology of Women Quarterly},
  volume={45},
  number={1},
  pages={106--125},
  year={2021},
  publisher={Sage Publications Sage CA: Los Angeles, CA}
}

@article{igarashi2022development,
  title={Development of the japanese version of the linguistic inquiry and word count dictionary 2015},
  author={Igarashi, Tasuku and Okuda, Shimpei and Sasahara, Kazutoshi},
  journal={Frontiers in psychology},
  volume={13},
  pages={841534},
  year={2022},
  publisher={Frontiers}
}

@article{iwanaga1998women,
  title={Women in Japanese politics: A comparative perspective},
  author={Iwanaga, Kazuki},
  year={1998},
  journal={Stockholm University, Center for Pacific Asia Studies}
}

@incollection{molony2018feminism,
  title={Feminism in Japan},
  author={Molony, Barbara},
  booktitle={Oxford Research Encyclopedia of Asian History},
  year={2018}
}

@inproceedings{ethayarajh2019understanding,
  title={Understanding Undesirable Word Embedding Associations},
  author={Ethayarajh, Kawin and Duvenaud, David and Hirst, Graeme},
  booktitle={Proceedings of the 57th Annual Meeting of the Association for Computational Linguistics},
  pages={1696--1705},
  year={2019}
}

@article{charlesworth2022historical,
  title={Historical representations of social groups across 200 years of word embeddings from Google Books},
  author={Charlesworth, Tessa ES and Caliskan, Aylin and Banaji, Mahzarin R},
  journal={Proceedings of the National Academy of Sciences},
  volume={119},
  number={28},
  pages={e2121798119},
  year={2022},
  publisher={National Acad Sciences}
}

@article{okuyama2021empowering,
  title   = {Empowering women through radio: Evidence from Occupied Japan},
  author  = {Okuyama, Yoko},
  journal = {Journal of Development Economics},
  volume  = {179},
  year    = {2026},
  pages   = {103620},
  issn    = {0304-3878},
  doi     = {10.1016/j.jdeveco.2025.103620},
  url     = {https://www.sciencedirect.com/science/article/pii/S0304387825001713}
}

@article{van2022equality,
  title={Equality of Men and Women in Article 24 of the Japanese Constitution (1947): The Role of Beate Sirota (1923-2012) and Beyond},
  author={van den Berg, Peter},
  journal={Osaka University Law Review},
  volume={69},
  number={1},
  pages={23--49},
  year={2022}
}

@article{gordon2021modern,
  title={A modern history of Japan from Tokugawa times to the present},
  author={Gordon, Andrew},
  year={2021},
  journal={Oxford University Press}
}

@article{garon1998molding,
  title={Molding Japanese minds: The state in everyday life},
  author={Garon, Sheldon},
  year={1998},
  journal={Princeton University Press}
}

@article{pennebaker2015development,
  title={The development and psychometric properties of LIWC2015},
  author={Pennebaker, James W and Boyd, Ryan L and Jordan, Kayla and Blackburn, Kate},
  year={2015}
}

@article{ward2019democratizing,
  title={Democratizing Japan: the allied occupation},
  author={Ward, Robert E and Sakamoto, Yoshikazu},
  year={2019},
  journal={University of Hawaii Press}
}

@article{pharr1981political,
  title={Political women in Japan: The search for a place in political life},
  author={Pharr, Susan J},
  year={1981},
  journal={Univ of California Press}
}

@article{moore2004partners,
  title={Partners for democracy: Crafting the new Japanese state under MacArthur},
  author={Moore, Ray A and Robinson, Donald L},
  year={2004},
  journal={Oxford University Press}
}

@article{matsui2024word,
  title={Word embedding for social sciences: An interdisciplinary survey},
  author={Matsui, Akira and Ferrara, Emilio},
  journal={PeerJ Computer Science},
  volume={10},
  pages={e2562},
  year={2024},
  publisher={PeerJ Inc.}
}

@article{michel2011quantitative,
  title={Quantitative analysis of culture using millions of digitized books},
  author={Michel, Jean-Baptiste and Shen, Yuan Kui and Aiden, Aviva Presser and Veres, Adrian and Gray, Matthew K and Google Books Team and Pickett, Joseph P and Hoiberg, Dale and Clancy, Dan and Norvig, Peter and others},
  journal={science},
  volume={331},
  number={6014},
  pages={176--182},
  year={2011},
  publisher={American Association for the Advancement of Science}
}

@misc{davies2008corpus,
  title={The corpus of contemporary American English: 450 million words, 1990-present},
  author={Davies, Mark},
  year={2008}
}

@article{tausczik2010psychological,
  title={The psychological meaning of words: LIWC and computerized text analysis methods},
  author={Tausczik, Yla R and Pennebaker, James W},
  journal={Journal of language and social psychology},
  volume={29},
  number={1},
  pages={24--54},
  year={2010},
  publisher={Sage Publications Sage CA: Los Angeles, CA}
}

@inproceedings{fast2016empath,
  title={Empath: Understanding topic signals in large-scale text},
  author={Fast, Ethan and Chen, Binbin and Bernstein, Michael S},
  booktitle={Proceedings of the 2016 CHI conference on human factors in computing systems},
  pages={4647--4657},
  year={2016}
}

@article{obana2003use,
  title={The use of kare/kanojo in Japanese society today},
  author={Obana, Yasuko},
  journal={New Zealand Journal of Asian Studies},
  volume={5},
  pages={139--155},
  year={2003}
}

@article{inoue2002gender,
  title={Gender, language, and modernity: Toward an effective history of Japanese women's language},
  author={Inoue, Miyako},
  journal={American Ethnologist},
  volume={29},
  number={2},
  pages={392--422},
  year={2002},
  publisher={Wiley Online Library}
}

@article{saito2016power,
  title={The power of translated literature in Japan: The introduction of new expressions through translation in the Meiji Era (1868--1912)},
  author={Saito, Mino},
  journal={Perspectives},
  volume={24},
  number={3},
  pages={417--430},
  year={2016},
  publisher={Taylor \& Francis}
}

@article{yamaguchi2019impediments,
  title={Impediments to the Advancement of Women in the Japanese Employment System: Theoretical Overview and the Purpose of This Book},
  author={Yamaguchi, Kazuo},
  journal={Gender Inequalities in the Japanese Workplace and Employment: Theories and Empirical Evidence},
  pages={1--45},
  year={2019},
  publisher={Springer}
}

@article{yamaguchi2016determinants,
  title={Determinants of the gender gap in the proportion of managers among white-collar regular workers in Japan},
  author={Yamaguchi, Kazuo},
  journal={Japan Labor Review},
  volume={13},
  number={3},
  pages={7--31},
  year={2016}
}

@incollection{yamaguchi2019causes,
  title={Causes and Effects of Gender Occupational Segregation: Overlooked Obstacles to Gender Equality},
  author={Yamaguchi, Kazuo},
  booktitle={Gender Inequalities in the Japanese Workplace and Employment: Theories and Empirical Evidence},
  pages={83--110},
  year={2019},
  publisher={Springer},
  address={Singapore}
}

@article{eto2020women,
  title={Women and Political Inequality in Japan: Gender Imbalanced Democracy},
  author={Eto, Mikiko},
  year={2020},
  journal={Routledge}
}

@article{ishiyama2019diachrony,
  title={Diachrony of Personal Pronouns in Japanese: A functional and cross-linguistic perspective},
  author={Ishiyama, Osamu},
  volume={344},
  year={2019},
  journal={John Benjamins Publishing Company}
}

@article{inohara2022jwsan,
  title={JWSAN: Japanese word similarity and association norm},
  author={Inohara, Keisuke and Utsumi, Akira},
  journal={Language Resources and Evaluation},
  volume={56},
  number={1},
  pages={109--137},
  year={2022},
  publisher={Springer}
}

@online{constitution,
  author       = {{2025 Ministry of Justice, Japan}},
  title        = {The Constitution of Japan},
  year         = {2025},
  url          = {https://www.japaneselawtranslation.go.jp/en/laws/view/174/tb},
  urldate      = {2025-9-11},
}

@online{laborstandard,
  author       = {{2025 Ministry of Justice, Japan}},
  title        = {Labor Standards Act},
  year         = {2025},
  url          = {https://www.japaneselawtranslation.go.jp/en/laws/view/3567/en},
  urldate      = {2025-9-11},
}

@online{oecd,
  author       = {{OECD}},
  title        = {Gender wage gap},
  year         = {2025},
  url          = {https://www.oecd.org/en/data/indicators/gender-wage-gap.html},
  urldate      = {2025-10-02},
}

@online{ngramdata,
  author       = {{GitHub}},
  title        = {NDL Ngram Data},
  year         = {2025},
  url          = {https://github.com/ndl-lab/ndlngramdata},
  urldate      = {2025-8-15},
}

@online{unidic,
  author       = {{UniDic}},
  year         = {2013},
  url          = {https://clrd.ninjal.ac.jp/unidic/back_number.html},
  urldate      = {2025-8-26}
}

@online{ngram2vec,
  author       = {{GitHub}},
  title       = {{Ngram2vec}},
  year         = {2018},
  url          = {https://github.com/zhezhaoa/ngram2vec},
  urldate      = {2025-9-01}
}

@online{occuclass,
  author       = {{Ministry of Internal Affairs and Communications}},
  title       = {{Japan Standard Occupational Classification}},
  year         = {1960},
  url          = {https://www.soumu.go.jp/toukei_toukatsu/index/seido/shokgyou/02toukatsu01_03000024.html},
  urldate      = {2025-10-02}
}

@online{nationalcensus,
  author       = {{e-Stat}},
  title       = {Japanese National Census},
  year         = {2025},
  url          = {https://www.e-stat.go.jp/stat-search/files?page=1&toukei=00200521},
  urldate      = {2025-10-06}
}

@article{kessler2019incentivized,
  title={Incentivized resume rating: Eliciting employer preferences without deception},
  author={Kessler, Judd B and Low, Corinne and Sullivan, Colin D},
  journal={American Economic Review},
  volume={109},
  number={11},
  pages={3713--3744},
  year={2019},
  publisher={American Economic Association 2014 Broadway, Suite 305, Nashville, TN 37203}
}

@article{moss2012science,
  title={Science faculty’s subtle gender biases favor male students},
  author={Moss-Racusin, Corinne A and Dovidio, John F and Brescoll, Victoria L and Graham, Mark J and Handelsman, Jo},
  journal={Proceedings of the national academy of sciences},
  volume={109},
  number={41},
  pages={16474--16479},
  year={2012},
  publisher={National Academy of Sciences}
}

@article{bowles2007social,
  title={Social incentives for gender differences in the propensity to initiate negotiations: Sometimes it does hurt to ask},
  author={Bowles, Hannah Riley and Babcock, Linda and Lai, Lei},
  journal={Organizational Behavior and human decision Processes},
  volume={103},
  number={1},
  pages={84--103},
  year={2007},
  publisher={Elsevier}
}

@article{buffington2016stem,
  title={STEM training and early career outcomes of female and male graduate students: Evidence from UMETRICS data linked to the 2010 census},
  author={Buffington, Catherine and Cerf, Benjamin and Jones, Christina and Weinberg, Bruce A},
  journal={American Economic Review},
  volume={106},
  number={5},
  pages={333--338},
  year={2016},
  publisher={American Economic Association 2014 Broadway, Suite 305, Nashville, TN 37203}
}

@article{amanatullah2010negotiating,
  title={Negotiating gender roles: Gender differences in assertive negotiating are mediated by women’s fear of backlash and attenuated when negotiating on behalf of others.},
  author={Amanatullah, Emily T and Morris, Michael W},
  journal={Journal of personality and social psychology},
  volume={98},
  number={2},
  pages={256},
  year={2010},
  publisher={American Psychological Association}
}

@article{schumann2010women,
  title={Why women apologize more than men: Gender differences in thresholds for perceiving offensive behavior},
  author={Schumann, Karina and Ross, Michael},
  journal={Psychological Science},
  volume={21},
  number={11},
  pages={1649--1655},
  year={2010},
  publisher={Sage Publications Sage CA: Los Angeles, CA}
}

@online{wordembeddingmodel,
  author       = {Shintaro Sakai},
  title        = {japanese-word-embeddings-100},
  year         = {2025},
  url          = {https://huggingface.co/shinsaka/japanese-word-embeddings-100/tree/main},
  urldate      = {2025-11-10},
}

@misc{NameDataset2021,
  author = {Philippe Remy},
  title = {Name Dataset},
  year = {2021},
  publisher = {GitHub},
  journal = {GitHub repository},
  howpublished = {\url{https://github.com/philipperemy/name-dataset}},
}

@online{enamdict,
  author       = {{GitHub}},
  title        = {enamdict Japanese Name Dataset},
  year         = {2025},
  url          = {https://github.com/rgamici/japanese-names},
  urldate      = {2026-1-30},
}

@online{name_origin,
  author       = {{GitHub}},
  title        = {name\_origin Japanese Name Dataset},
  year         = {2025},
  url          = {https://github.com/shuheilocale/japanese-personal-name-dataset},
  urldate      = {2026-1-30},
}

@misc{pham2023gendec,
  title        = {Gendec: A Machine Learning-based Framework for Gender Detection from Japanese Names},
  author       = {Pham, Duong Tien and Nguyen, Luan Thanh},
  year         = {2023},
  eprint       = {2311.11001},
  archivePrefix= {arXiv},
  primaryClass = {cs.CL},
  url          = {https://huggingface.co/datasets/tarudesu/gendec-dataset}
}

@book{north1990institutions,
  title={Institutions, institutional change and economic performance},
  author={North, Douglass C},
  year={1990},
  publisher={Cambridge University Press},
  address={Cambridge}
}

@article{williamson2000new,
  title={The new institutional economics: taking stock, looking ahead},
  author={Williamson, Oliver E},
  journal={Journal of economic literature},
  volume={38},
  number={3},
  pages={595--613},
  year={2000},
  publisher={American Economic Association}
}

@article{roland2004understanding,
  title={Understanding institutional change: Fast-moving and slow-moving institutions},
  author={Roland, G{\'e}rard},
  journal={Studies in comparative international development},
  volume={38},
  number={4},
  pages={109--131},
  year={2004},
  publisher={Springer}
}

@article{north1994economic,
  title={Economic performance through time},
  author={North, Douglass C},
  journal={The American economic review},
  volume={84},
  number={3},
  pages={359--368},
  year={1994},
  publisher={JSTOR}
}

@article{williamson2009informal,
  title={Informal institutions rule: institutional arrangements and economic performance},
  author={Williamson, Claudia R},
  journal={Public Choice},
  volume={139},
  number={3},
  pages={371--387},
  year={2009},
  publisher={Springer}
}

@article{england2010gender,
  title={The gender revolution: Uneven and stalled},
  author={England, Paula},
  journal={Gender \& society},
  volume={24},
  number={2},
  pages={149--166},
  year={2010},
  publisher={Sage Publications Sage CA: Los Angeles, CA}
}

@article{goldin2014grand,
  title={A grand gender convergence: Its last chapter},
  author={Goldin, Claudia},
  journal={American economic review},
  volume={104},
  number={4},
  pages={1091--1119},
  year={2014},
  publisher={American Economic Association 2014 Broadway, Suite 305, Nashville, TN 37203}
}

@article{sayer2011she,
  title={She left, he left: How employment and satisfaction affect women’s and men’s decisions to leave marriages},
  author={Sayer, Liana C and England, Paula and Allison, Paul D and Kangas, Nicole},
  journal={American Journal of Sociology},
  volume={116},
  number={6},
  pages={1982--2018},
  year={2011},
  publisher={University of Chicago Press Chicago, IL}
}
\end{document}